\documentclass[useAMS,usenatbib]{mn2e}

\usepackage{graphicx}
\let\Newcommand=\newcommand
\def\newcommand{\providecommand}
\usepackage[pdftitle={Galaxy Cluster Masses Without Non-Baryonic Dark Matter}, pdfauthor={J. R. Brownstein and J. W. Moffat},
pdfsubject={\tt astrophysics, Galaxies, Clusters, Gravitation, Quantum Gravity, RG Flow}, pdfkeywords={rotation curves, galaxies, quantum gravity, mond, dark matter}, bookmarks, bookmarksopen, 
colorlinks=true, linkcolor=blue, citecolor=blue, anchorcolor=blue, urlcolor=blue]{hyperref}
\let\newcommand=\Newcommand

\newcommand{\etal}{{\em et al.}\ }
\newcommand\colhead[1]{\multicolumn{1}{c}{#1}}%
\newcommand\aj{AJ}%
\newcommand\araa{ARA\&A}%
\newcommand\apj{ApJ}%
\newcommand\aap{A\&A}%
\newcommand\jcap{J. Cosmology Astropart. Phys.}%
\newcommand\mnras{MNRAS}%
\newcommand\prd{Phys.~Rev.~D}%
\title[Galaxy Cluster Masses Without Non-Baryonic Dark Matter]{Galaxy Cluster Masses Without Non-Baryonic Dark Matter}
\author[J. R. Brownstein and J. W. Moffat]{J.\,R. Brownstein\thanks{\href{mailto:jbrownstein@perimeterinsitute.ca}{\tt jbrownstein@perimeterinsitute.ca}} and
J.\,W. Moffat\thanks{\href{mailto:john.moffat@utoronto.ca}{\tt john.moffat@utoronto.ca}}\\
The Perimeter Institute for Theoretical Physics, Waterloo, Ontario, N2J 2W9, Canada, and\\
Department of Physics, University of Waterloo, Waterloo, Ontario N2Y 2L5, Canada}
\begin{document}

\date{Submitted 2005 July 8.}

\pagerange{\pageref{firstpage}--\pageref{lastpage}} \pubyear{2005}

\maketitle

\label{firstpage}

\begin{abstract}
We apply the modified acceleration law obtained from Einstein gravity coupled to a massive skew
symmetric field, $F_{\mu\nu\lambda}$, to the problem of explaining X-ray galaxy cluster masses without exotic dark
matter.
 Utilizing X-ray observations to fit the gas mass profile and temperature profile of the hot intracluster medium (ICM)
with King ``$\beta$-models'', we show that the dynamical masses of the galaxy clusters resulting from our modified
acceleration law fit the cluster gas masses for our sample of 106 clusters without the need of introducing a
non-baryonic dark matter component.  We are further able to show for our sample of 106 clusters that the distribution
of gas in the ICM as a function of radial distance is well fit by the dynamical mass distribution arising
from our modified acceleration law without any additional dark matter component.  In previous work, we applied this
theory to galaxy rotation curves and demonstrated good fits to our sample of 101 LSB, HSB and dwarf galaxies including
58 galaxies that were fit photometrically with the single parameter $(M/L)_{stars}$.  The results there were
qualitatively similar to those obtained using Milgrom's phenomenological MOND model, although the determined galaxy masses
were quantitatively different and MOND does not show a return to Keplerian behavior at extragalactic distances.  The
results here are compared to those obtained using Milgrom's phenomenological MOND model which does not fit the X-ray
galaxy cluster masses unless an auxiliary dark matter component is included.
\end{abstract}


\section{Introduction}

The question of galaxy rotation curves and cluster masses has been known to require some form of energy density that 
makes its presence felt only by its gravitational effects since \citet{zwi33} analyzed the velocity dispersion for the
Coma cluster.  By 1959 there were eight galaxy rotation curves available from radio observations demonstrating a ``flat
rotation velocity''.  In \citet{bro05} -- our study of galaxy rotation curves without non-baryonic dark matter -- we
demonstrated good fits to our sample of 101 LSB, HSB and dwarf galaxies including 58 galaxies that were fit
photometrically with the single parameter $(M/L)_{stars}$ (29 $B-$band and 29 $K-$band).   
As a follow-up, we apply here the same framework to the question of X-ray galaxy cluster
masses~\citep{mof05a,mof05b,bro05}.

In this framework, we apply a generalization of Einstein's general relativity (GR) based on a pseudo-Riemannian metric
tensor and a skew symmetric rank three tensor field $F_{\mu\nu\lambda}$, called metric-skew-tensor-gravity (MSTG).  A
renormalization group (RG) framework~\citep{reu04a,reu04b} for MSTG was developed to describe the running of the
effective gravitational coupling constant $G$, and the effective coupling constant $\gamma_c$ that measures the strength
of the coupling of the $F_{\mu\nu\lambda}$ field to matter. A momentum cutoff identification $k=k(x)$ associates the RG
scales to points in spacetime. For the static, spherically symmetric solution, the RG flow equations allow a running
with momentum $k$ and proper length $\ell(r)=1/k$ for the effective Newtonian coupling constant $G=G(r)$, the coupling
constant $\gamma_c=\gamma_c(r)$, and the effective mass of the skew field $\mu=\mu(r)$ where $r$ denotes the radial
coordinate. The form of $G(r)$ as a function of $r$, obtained from the modified Newtonian acceleration law, leads to
agreement with solar system observations, terrestrial gravitational experiments and the binary pulsar PSR 1913+16
observations, while strong renormalization effects in the infrared regime at large distances lead to fits to galaxy
rotation curves and X-ray cluster masses. In \citet{mof05b}, a scalar-tensor-vector gravity theory is formulated, which 
yields the same weak field modified acceleration law as in MSTG \citet{mof05a}, and equations that describe an effective
running of $G$, $\gamma_c$ and $\mu$ with space and time.

A fit to the bulk properties (the gas mass) of 106 X-ray galaxy clusters is achieved without exotic dark matter by a
best fit result using a nonlinear least-squares fitting routine including estimated errors for the one free parameter in
the RG flow equations, $M_{0}$, which fully describes the large distance renormalization of Newton's constant,
$G_{\infty}$.  The running of Newton's constant, $G(r)$, is fully constrained by choosing the remaining free parameter in
the RG flow equations, $r_{0}$, to take on a value proportional to the scale of the galaxy cluster (for small clusters
with galactic scale ICM gas masses) or to take on a particular constant value for the majority of the regular
size galaxy clusters (where the ICM gas mass dominates over the masses of the individual galaxies).  Thus, the
individual cluster mass profiles as functions of radial position throughout the range of X-ray observations are parameter
free predictions; and are consistent with X-ray observed gas mass profiles without exotic dark matter.

The galaxy cluster's bulk properties fit and the specific fit to the individual clusters are compared to those obtained
using Milgrom's phenomenological MOND model which does not fit the X-ray galaxy cluster masses unless an auxiliary dark
matter component is included~\citep{san03}.  \citet{the88} were able to account for the MOND discrepancy between the X-ray
observationally determined gas mass and the dynamical mass of the Coma cluster by increasing the MOND acceleration by
a factor of four greater than that used to fit the galaxy rotation curves.  \citet{agu01} present
evidence from the central 200 kpc of three clusters implying that MOND is inconsistent with the observed temperature
gradient which inflates the discrepancy in the MOND acceleration to a factor of $\sim$ 10. More
recently, \citet{poi05b} use X-ray data from the {\sc XMM-Newton} satellite for eight clusters of varying temperature
and masses to place constraints on the use of the MOND phenomenology.  Without
treating the MOND acceleration as a free parameter as opposed to a universal constant, MOND predicts dynamical masses
greatly in excess of the X-ray observations -- necessitating the ad hoc addition of dark matter to explain away the
missing mass.  We are able to show that there is no missing mass when applying the MSTG acceleration law to galaxy
clusters.

\section{Isotropic Isothermal Model}

Recent observations from the {\sc XMM-Newton} satellite suggest that the intracluster medium (ICM) is very nearly isothermal
inside the region defined by the X-ray emission with temperatures ranging from $\approx$ 1--15 keV (or $10^{7}$ -- $2 \times
10^{8}$ K) for different clusters~\citep{arn01b}.  The combination of the observed density profile, $n_{e}(r)$, and the
temperature profile, $T(r)$, obtained from X-ray observations
of the galaxy cluster leads to a pressure profile, $P(r)$, which directly leads to a mass profile, $M(r)$, by assuming the
gas is in nearly hydrostatic equilibrium with the gravitational potential of the galaxy cluster. Within a
few core radii, the distribution of gas within a galaxy cluster may be fit by a King ``$\beta$-model''.  The
observed surface brightness of the X-ray cluster can be fit to a radial distribution profile~\citep{cha60,kin66}:
\begin{equation}
\label{betaModel} I(r)= I_{0}\left[  1+\left(\frac{r}{r_{c}}\right)^{2}\right]^{-3 \beta + 1/2},
\end{equation}
resulting in best fit parameters, $\beta$ and $r_{c}$. A deprojection of the $\beta$-model of equation
(\ref{betaModel}) assuming a nearly isothermal gas sphere then results in a physical gas density
distribution~\citep{cav76}:
\begin{equation}
\label{betaRhoModel} \rho(r)= \rho_{0}\left[  1+\left(\frac{r}{r_{c}}\right)^{2}\right]^{-3 \beta /2},
\end{equation}
where $\rho(r)$ is the ICM mass density profile.  Provided the number density, $n$,  traces the actual mass, we may
assume that $n(r) \propto \rho(r)$, which according to \citet{rei01} is explicitly
\begin{equation}
\label{numberMass} \rho_{\rm gas} \approx 1.17 n_{e} m_{p},
\end{equation}
and rewrite equation (\ref{betaRhoModel}) 
\begin{equation}
\label{betaNumberModel} n_{e}(r)= n_{0}\left[  1+\left(\frac{r}{r_{c}}\right)^{2}\right]^{-3 \beta /2}.
\end{equation}
For a spherical system in hydrostatic equilibrium, the structure equation can be derived from the collisionless Boltzmann
equation
\begin{equation}
\label{CBE} \frac{d}{dr}(\rho(r) \sigma_{r}^{2}) + \frac{2\rho(r)}{r}\left(\sigma_{r}^{2} - \sigma_{\theta,\phi}^{2}\right) =
-\rho(r) \frac{d\Phi(r)}{dr},
\end{equation}
where $\Phi(r)$ is the gravitational potential for a point source, $\sigma_{r}$ and $\sigma_{\theta,\phi}$ are mass-weighted
velocity dispersions in the radial ($r$) and tangential ($\theta, \phi$) directions, respectively.  For an isotropic
system,
\begin{equation}
\label{isotropicSystem}
\sigma_{r} = \sigma_{\theta,\phi}.
\end{equation}
The pressure profile, $P(r)$, can be related to these quantities by
\begin{equation}
\label{pressureProfile}
P(r) = \sigma_{r}^{2} \rho(r).
\end{equation}
Combining equations (\ref{CBE}), (\ref{isotropicSystem}) and (\ref{pressureProfile}), the result for the isotropic
sphere is
\begin{equation}
\label{isotropicSphere}
\frac {dP(r)}{dr} = -\rho(r) \frac{d\Phi(r)}{dr}.
\end{equation}
For a gas sphere with temperature profile, $T(r)$, the velocity dispersion becomes
\begin{equation}
\label{velocityDispersion}
\sigma_{r}^{2} = \frac{kT(r)}{\mu m_{p}},
\end{equation}
where $k$ is Boltzmann's constant, $\mu \approx 0.609$ is the mean atomic weight and $m_{p}$ is the proton mass.  We
may now substitute equations (\ref{pressureProfile}) and (\ref{velocityDispersion}) into equation (\ref{isotropicSphere}) to obtain
\begin{equation}
\label{isotropicSphere2}
\frac {d}{dr}\left(\frac{kT(r)}{\mu m_{p}} \rho(r)\right) = -\rho(r) \frac{d\Phi(r)}{dr}.
\end{equation}
Performing the differentiation on the left hand side of equation (\ref{isotropicSphere}), we may solve for the
gravitational acceleration:
\begin{eqnarray}
\nonumber
a(r) & \equiv  & - \frac{d\Phi(r)}{dr} \\
\label{accelerationProfile}
& = & \frac{kT(r)}{\mu m_{p} r} \left[ \frac{d \ln(\rho(r))}{d \ln(r)} + \frac{d \ln(T(r))}{d \ln(r)}\right].
\end{eqnarray}
For the isothermal isotropic gas sphere, the temperature derivative on the right-hand side of equation
(\ref{accelerationProfile}) vanishes and the remaining derivative can be evaluated using the $\beta$-model of equation (\ref{betaRhoModel}):
\begin{equation}
\label{isothermalAccelerationProfile}
a(r) = -\frac{3\beta kT}{\mu m_{p}} \left(\frac{r}{r^{2}+r_{c}^{2}}\right).
\end{equation}

\section{Mass Profiles}
Given an isotropic density distribution, $\rho(r)$, the associated mass profile is
\begin{equation}
\label{massProfile}
M(r)  = 4\pi\int_{0}^{r} \rho(r') r'^{2} dr',
\end{equation}
where $M(r)$ is the total mass contained within a sphere of radius r.  For the $\beta$-model of equation
(\ref{betaRhoModel}), we may approximate the integral of equation (\ref{massProfile}) for $r \gg r_{c}$ and
$\beta<1$~\citep{rei01}:
\begin{equation}
\label{totalMassApproximation}
M(r)  \approx \frac{4\pi\rho_{0}r_{c}^{3}}{3(1-\beta)}
\left(\frac{r}{r_{c}}\right)^{3(1-\beta)} \longleftarrow \{r \gg r_{c}\ \mbox{and}\ \beta < 1\}.
\end{equation}
This result clearly diverges in the limit as $r \rightarrow \infty$; but galaxy clusters are observed to have finite 
spatial extent.  This allows an approximate determination of the total mass of the galaxy cluster by first solving
equation (\ref{betaRhoModel})
for the position, $r_{\rm out}$, at which the density, $\rho(r_{\rm out})$, drops to $\approx 10^{-28}\,\mbox{g/cm}^{3}$, or 250
times the mean cosmological density of baryons:
\begin{equation}
\label{rout}
r_{\rm out} = r_{c} \sqrt{\left(\frac{\rho_{0}}{10^{-28}\,\mbox{g/cm}^{3}}\right)^{2/3\beta}-1}.
\end{equation}
Then, provided $\beta < 1$,
\begin{equation}
\label{totalGasMass}
M_{\rm gas} \approx \frac{4\pi\rho_{0}r_{c}^{3}}{3(1-\beta)}\left(\frac{r_{\rm out}}{r_{c}}\right)^{3(1-\beta)}.
\end{equation}

The dynamical mass in Newton's theory of gravitation can be obtained as a function of radial position by replacing the
gravitational acceleration with Newton's Law:
\begin{equation}
\label{newtonsLaw}
a_{\rm N}(r) = - \frac{G_{0} M(r)}{r^{2}},
\end{equation}
where $G_0$ is Newton's ``bare'' gravitational constant so that equation (\ref{accelerationProfile}) can be rewritten as
\begin{equation}
\label{newtonsMass}
M_{\rm N}(r) = - \frac{r}{G_{0}}\frac{kT}{\mu m_{p}} \left[ \frac{d \ln(\rho(r))}{d \ln(r)} + \frac{d \ln(T(r))}{d
\ln(r)}\right],
\end{equation}
and the isothermal $\beta$-model result of equation (\ref{isothermalAccelerationProfile}) can be
rewritten as
\begin{equation}
\label{isothermalNewtonsMass}
M_{\rm N}(r) = \frac{3\beta kT}{\mu m_{p}G_{0}} \left(\frac{r^{3}}{r^{2}+r_{c}^{2}}\right).
\end{equation}

Similarly, the dynamical mass in MSTG can be obtained as a function of radial position by substituting the MSTG gravitational
acceleration law~\citep{mof05a,mof05b,bro05}:
\begin{equation}
\label{runG} a(r)=-\frac{G(r)M}{r^2},
\end{equation}
so that our result for the isothermal $\beta$-model becomes
\begin{equation}
\label{mstgMassProfile}
M_{\rm MSTG}(r) = \frac{3\beta kT}{\mu m_{p}G(r)} \left(\frac{r^{3}}{r^{2}+r_{c}^{2}}\right).
\end{equation}
We can express this result as a scaled version of equation (\ref{newtonsMass}) or the isothermal case of equation
(\ref{isothermalNewtonsMass}):
\begin{eqnarray}
\nonumber M_{\rm MSTG}(r)  &=& \frac{G_{0}}{G(r)} M_{\rm N}(r)\\
\nonumber &=& \biggl\{1+\sqrt\frac{M_0}{M_{\rm MSTG}(r)}\biggl[1-\\
 \label{mstgMassScaled}&\phantom{=}& \quad\exp(-r/r_0)\biggl(1+\frac{r}{r_0}\biggr)
\biggr]\biggr\}^{-1} M_{\rm N}(r),
\end{eqnarray}
where we have made explicit the form of the running of G(r) as in \citet{mof05a,mof05b,bro05}:
\begin{equation}
\label{runningG} G(r) = \biggl\{1+\sqrt\frac{M_0}{M_{\rm MSTG}(r)}\biggl[1-\exp(-r/r_0)\biggl(1+\frac{r}{r_0}\biggr)
\biggr]\biggr\}.
\end{equation}

In the limit of large $r$, we show in \citet{mof05a,mof05b,bro05} that
\begin{equation}
\label{Ginfinity} G_{\infty}\equiv \lim_{r \gg r_{0}} G(r) = G_{0} \left\{1 +
\sqrt{\frac{M_{0}}{M_{\rm gas}}}\right\},
\end{equation}
and the total mass of the cluster in MSTG can be computed by taking equations (\ref{mstgMassScaled}) to the same limit:
\begin{eqnarray}
\nonumber M_{\rm MSTG} & = & \frac{G_{0}}{G_{\infty}} M_{\rm N}\\
\label{mstgMass} & = & \left\{1 +\sqrt{\frac{M_{0}}{M_{\rm MSTG}}}\right\}^{-1} M_{\rm N}.
\end{eqnarray}

It is a simple matter to solve equations (\ref{mstgMassScaled}) and (\ref{mstgMass}) explicitly for $M_{\rm MSTG}(r)$
and $M_{\rm MSTG}$, respectively, by squaring both sides and subsequently applying the canonical solution to the quadratic equation.

The derivation in Milgrom's phenomenological MOND model~\citep{mil83,san02} follows the same procedure, but utilizes the MOND
gravitational acceleration law, described by
\begin{equation}
\label{milgromacc} a \mu\biggl(\frac{a}{{a_0}_{\rm Milgrom}}\biggr)=a_{\rm Newton},
\end{equation}
where $\mu(x)$ is a function
that interpolates between the Newtonian regime, $\mu(x)=1$, when $x\gg 1$ and the MOND
regime, $\mu(x)=x$, when $x\ll 1$. The function and critical acceleration normally used for galaxy and cluster fitting are, respectively, 
\begin{eqnarray}
\mu(x) & = &\frac{x}{\sqrt{1+x^2}},\\
\label{a0Mond} {a_0}_{\rm Milgrom} & = & 1.2\times 10^{-8}\,{\rm cm}/{\rm s}^2.
\end{eqnarray}
Applying equation (\ref{milgromacc}) to either equation (\ref{accelerationProfile}) or the isothermal case of equation
(\ref{isothermalAccelerationProfile}) yields the MOND dynamical mass in terms of the Newtonian dynamical mass of
equation (\ref{newtonsMass}) or the isothermal case of equation (\ref{isothermalNewtonsMass}):
\begin{equation}
\label{mondMass} M_{\rm MOND}(r) = \frac{M_{\rm N}(r)}{\sqrt{1+\left({a_0}_{\rm Milgrom} / \frac{G_{0} M_{\rm 
MOND}(r)}{r^{2}}\right)^{2}}}.
\end{equation} 
It is a simple matter to solve equation (\ref{mondMass}) explicitly for
$M_{\rm MOND}(r)$ by squaring both sides and subsequently applying the canonical solution to the quadratic equation.

Unlike the Newtonian and MSTG isothermal spheres whose densities fall off as $1/r^{2}$ in the limit of large $r$,
\citet{mil84} showed that the MOND isothermal spheres have densities that fall of as $r^{-\alpha}$ where $\alpha
\approx 4$ at large radii.  In the Newtonian and MSTG cases, the behavior is Keplerian at large radii
and so the isothermal spheres have masses which diverge with $r$~\citep{cha60}.  Thus according to equations
(\ref{newtonsLaw}) and (\ref{runG}) the acceleration at large radii for Newtonian and MSTG isothermal spheres goes as
$1/r$.  Conversely, the MOND isothermal spheres have convergent masses regardless of the spatial extent~\cite{san03}:
\begin{equation}
\label{totalMondMass} M_{\rm MOND} \approx \frac{16}{G_{0} {a_0}_{\rm Milgrom}} \left( \frac{k T}{\mu m_{p}}\right)^{2}.
\end{equation} 
The MOND acceleration law regularizes the integral of equation (\ref{massProfile}) whereas a cutoff is necessary in
the Newtonian and MSTG case, corresponding to the finite spatial extent of galaxy clusters as shown in equation
(\ref{totalGasMass}).  However, as clearly shown in \citet{san03}, the X-ray surface brightness distribution of these
MOND isothermal spheres provides a poor representation of the observed surface brightness distribution.  This distinction
will become apparent in the section ahead where we study the MSTG and MOND mass profiles of the Coma cluster.

\section{Running of Newton's Constant}
We have adopted the compilation of \citet{rei01, rei02} as our sample.  The relevant cluster properties are listed in
Table~\ref{clusterProperties} arranged as follows:  Column (1) lists the cluster name truncated to 8 characters.  Column
(2) is the observed X-ray temperature.  Column (3) lists the ICM central mass density, $\rho_{0}$, of the ICM in units of
$10^{-25}\,\mbox{g/cm}^{3}$.  Column (4) is the $\beta$ parameter.  Column (5) is the core radius parameter, $r_{c}$, in
units of kpc assuming $H_{0} = 0.71_{-0.03}^{+0.04}$~\citep{eid04}.  Column (6) is the position, $r_{\rm out}$ at which
the density, $\rho(r_{\rm out})$, drops to $\approx 10^{-28}\,\mbox{g/cm}^{3}$, or 250 times the mean cosmological
density of baryons.  Column (7) through (9) lists the ICM gas mass, $M_{\rm gas}$, the Newtonian dynamic mass,
$M_{\rm N}$, and the MSTG dynamic mass, $M_{\rm MSTG}$, respectively, each integrated to $r_{\rm out}$. Column (10)
lists the ``convergent'' MOND dynamic mass.

In order to calculate the MSTG dynamic mass we first need to phenomenologically determine the
running of the parameters, $M_{0}$, and $r_{0}$ -- this describes the running of Newton's
constant at the scale of clusters according to equation (\ref{runningG}).  However, unlike the case of the galaxy
rotation curves where a satisfactory fit to LSB and HSB galaxy data is obtained with the
parameters~\citep{bro05}
\begin{equation}
\label{galaxyParameters} M_0=9.60\times 10^{11}M_{\sun},\quad r_0=13.92\,{\rm kpc},
\end{equation}
we found that a better fit was attained by dropping the simplifying assumption that $M_{0}$ is constant
across clusters.  In fact, we were able to account for all dwarf galaxies smaller than 12 kpc in \citet{bro05} by
allowing $M_{0}$ to scale down to:
\begin{equation}
\label{dwarfParameters} M_0=2.40\times 10^{11}M_{\sun},\quad r_0=6.96\,{\rm kpc}.
\end{equation}
The three schemes we attempted to prescribe for the scale variation of $M_{0}$ for clusters are as follows:
\begin{enumerate}
\item $M_{0} \propto M_{\rm gas}^{n}$, where $M_{\rm gas}$ is the total ICM gas mass integrated to
$r_{\rm out}$,
\item $M_{0} \equiv \rm{constant}$,
\item $G_{\infty} \equiv \rm{constant}$.
\end{enumerate}
It is clear that case (ii) is the limit of case (i) taking $n \rightarrow 0$ and also that case (iii) is the limit of
case (i) taking $n \rightarrow 1$.  By plotting $M_{\rm N}$ against $M_{\rm gas}$ and then applying equation
(\ref{mstgMass}), we were able to constrain the parameter, $M_{0}$, using a nonlinear least-squares fitting
routine including estimated errors .  The fits for each of the three above schemes are shown in Figure~\ref{dynamicMass}, with the results as follows:
\begin{enumerate}
\item $M_{0} = (58.8 \pm 4.0) \times 10^{14} M_{\sun} \left(\frac{M_{\rm gas}}{10^{14} M_{\sun}}\right)^{0.39 \pm
0.09}$,
\item $M_{0} = (59.4 \pm 4.3) \times 10^{14} M_{\sun}$,
\item $G_{\infty} = 7.84\pm0.27$.
\end{enumerate}
The quality of each of these prescriptions for $M_{0}$ is seen in their respective plots of Figure~\ref{dynamicMass}
where the MSTG mass is plotted against the ICM gas mass.  Clearly, the plot corresponding to case (i) is the best
scheme since the slope of unity best describes the
bulk properties of the clusters.  Indeed, the overall least sum of squares best fit is the scheme corresponding to
case (i) with the other two acting as limiting cases.   Meanwhile, both the MOND ``convergent mass'' and the Newtonian
dynamic mass show a discrepancy with the ICM gas mass~\citep{san03}.

In the galaxy rotation curves, we scaled $r_{0}$ down from equation (\ref{galaxyParameters}) by a factor of 2 to account
for the Dwarf galaxies~\citep{bro05}.
 For all but the smallest galaxy clusters where the mass of the ICM dominates over the mass of the individual galaxies,
scaling $r_{0}$ up from equation (\ref{galaxyParameters}) by a factor of 10 leads to satisfactory
fits, whereas for the smallest galaxy clusters where the mass of the ICM is of the order of galactic masses -- with
$r_{\rm out}
\leq 650\,\mbox{kpc}$ -- scaling $r_{0}$ to $r_{\rm out}/10$ leads to satisfactory fits.   Thus our prescription for the
running of Newton's constant (RG flow) for galaxy clusters is fully constrained:
\begin{eqnarray}
\nonumber M_{0} &=& 58.8 \times 10^{14} M_{\sun} \left(\frac{M_{\rm gas}}{10^{14} M_{\sun}}\right)^{0.39},
\end{eqnarray}
\begin{eqnarray}
\label{clusterParameters} r_0 =& r_{\rm out}/10, &r_{\rm out} \leq 650\,\mbox{kpc},\\
\nonumber r_0 =& 139.2\,{\rm kpc}, &r_{\rm out} > 650\,\mbox{kpc}.
\end{eqnarray}

\section{Individual Clusters} We now turn to the individual mass profiles for each cluster.  Since the MSTG
prediction for the mass profile has no free parameters upon enforcing equations (\ref{clusterParameters}), we need not
perform any nonlinear least-squares fitting for the individual mass profiles as shown in Figure~\ref{clusters}.   The
Coma cluster is shown first (with errorbands); and the remaining 105 clusters follow as thumbnails (without errorbands for
clarity).

The $\beta$-model is not always a good fit to the data, however, and so the mass profiles shown in Figure~\ref{clusters}
have limited reliability and the results of Table~\ref{clusterProperties} show errors ranging from a few percent to in
some cases 100\%.  \citet{ota02} show that some clusters observed with {\sc ROSAT} and {\sc ASCA} are
expressed better by the double $\beta$-model with a second core; while \citet{poi05a} show that clusters observed
with {\sc XMM-Newton} are well described by an {\sc NFW}-type profile~\citep{nav97} showing strong agreement with CDM
simulations.  While these more sophisticated mass profile models have their advantages, the simplest 
isothermal isotropic $\beta$-model based upon hydrostatic equilibrium has the fewest parameters and we have used it to
show that it is possible and meaningful to fit the X-ray galaxy cluster data without the need for exotic dark matter.

There are also controversial measurements concerning the presence of temperature gradients in galaxy
clusters~\citep{rei02}.  {\sc XMM-Newton} is expected to probe the isothermal properties of the galaxy clusters.
Early results such as the \citet{arn01a} study of the {\sc XMM-Newton} observation of the Coma galaxy cluster
and the \citet{arn01b} study of the {\sc XMM-Newton} radial temperature profiles support the isothermal assumption
deviating only near the cluster center where the observed X-ray temperatures decrease.

In those cases where the uncertainty in the $\beta$-model parameters are sufficiently small (and therefore the $\beta$-model
provides a suitable fit), the MSTG mass profile traces the ICM gas mass profile down to sufficiently small radii and
only deviates where the assumption that the ICM gas mass is isothermal is no longer valid.  At sufficiently small radii,
the observed X-ray temperatures decrease and thus the MSTG prediction based on an isothermal sphere is too large,
confirmed by Figure~\ref{clusters}.

\section{Conclusions}

A gravity theory consisting of a metric-skew-tensor action (MSTG) that
leads to the modified acceleration law~\citep{mof05a,mof05b,bro05} can be fitted to a large class of X-ray galaxy
cluster mass profiles.  The same acceleration law can also be applied to fit a large class of galaxy
rotation curves as in \citet{bro05}  -- which achieved one parameter fits for the case of the photometric observations
with a single parameter $(M/L)_{\rm stars}$; and also parametric fits which required no simplifying assumption on the
constancy of (M/L) throughout the galaxy.  Whereas MOND provides a good phenomenological fit to galaxy rotation curves,
only MSTG fits both galaxy rotation curves and the X-ray galaxy cluster mass profiles without the need of introducing a
non-baryonic dark matter component.

In addition, whereas MOND necessitates the asymptotic ``flat rotation'' velocity out to infinity, MSTG only gives the
appearance of a ``flat rotation'' velocity within the confines of the galaxy and returns to the familiar Newton-Kepler
form at large distances.  Using the Sloan Digital Sky Survey (SDSS), \citet{pra03} have studied the velocities of
satellites orbiting isolated galaxies. They detected approximately 3000 satellites, and they found that the
line-of-sight velocity dispersion of satellites declines with distance to the primary. The velocity was observed to
decline to a distance of $\sim 350$ kpc for the available data. This result contradicts the constant velocity prediction
equation of MOND, but is consistent with the MSTG prediction.

The newest results from {\sc XMM-Newton} reveal that the mass profiles of X-ray clusters show a steady rise out to the
limit of X-ray observations~\citep{poi05a} and do not show the convergent behavior of MOND isothermal spheres. 
However, the {\sc XMM-Newton} results are consistent with the MSTG isothermal sphere whose densities fall off as
$1/r^{2}$ without unseen non-baryonic dark matter.  Further results from {\sc XMM-Newton} may
reveal that the temperature profiles of X-ray clusters are not everywhere isothermal; and observed gradients may add a
greater understanding of the physics of X-ray clusters of galaxies and the law of gravitation that best describes these
largest of virialized objects.

\section*{Acknowledgments}

This work was supported by the Natural Sciences and Engineering Research Council of Canada.  We thank Thomas
H.\,Reiprich for supplying the data for our galaxy cluster sample via \href{http://www.astro.virginia.edu/~thr4f/act/gcs/}{electronic
table}\footnote{\url{http://www.astro.virginia.edu/~thr4f/act/gcs/}}.

\clearpage

\setcounter{figure}{0}
\begin{figure*}
\begin{center}
\begin{tabular}{cc}
\includegraphics[width=82.5mm]{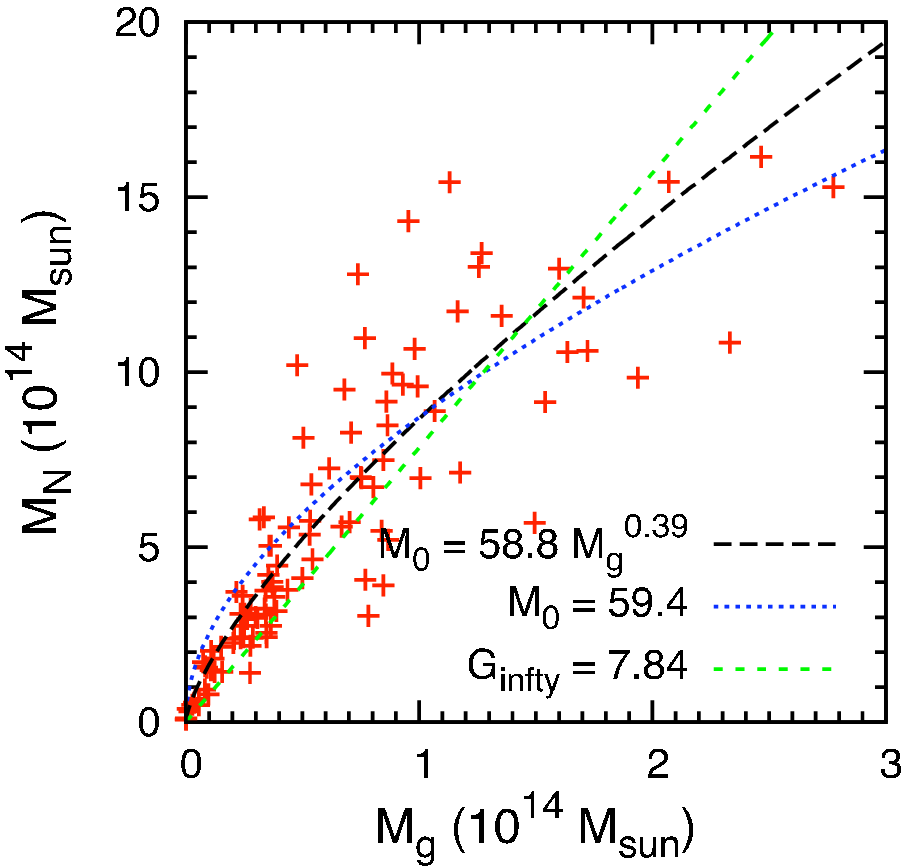} & \includegraphics[width=82.5mm]{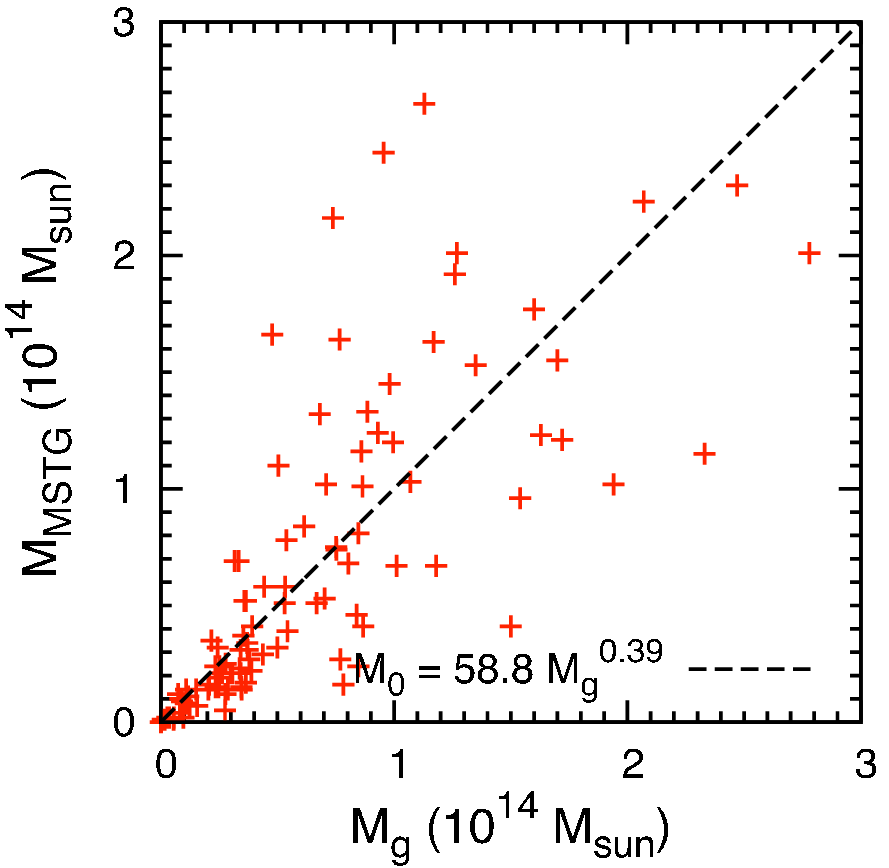} \\
\includegraphics[width=82.5mm]{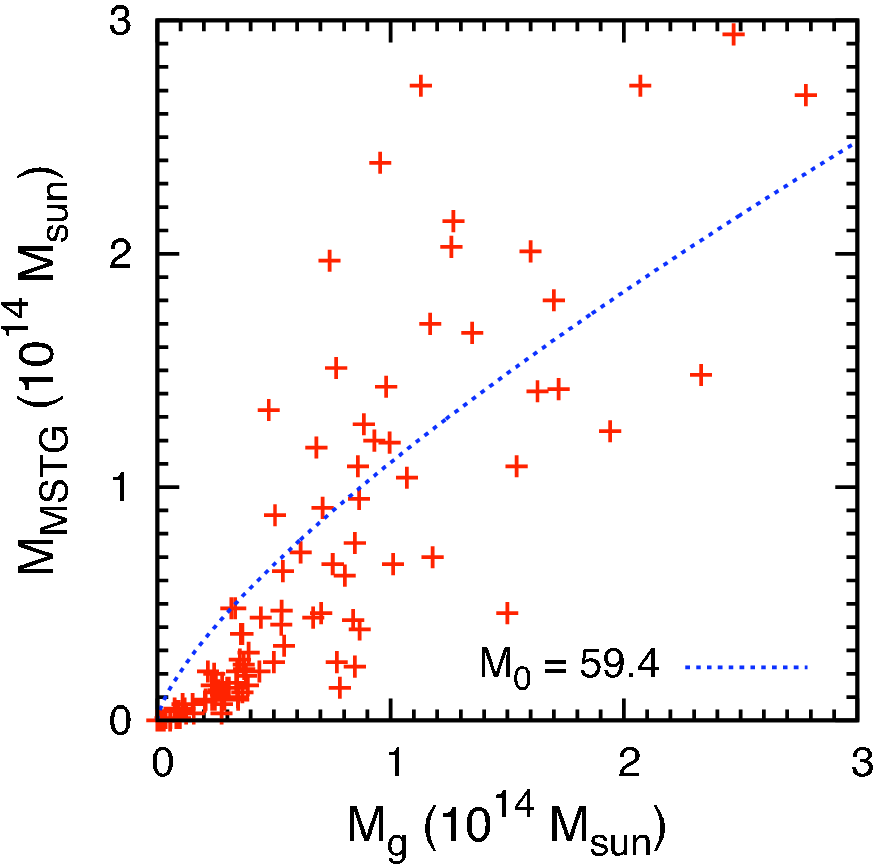} & \includegraphics[width=82.5mm]{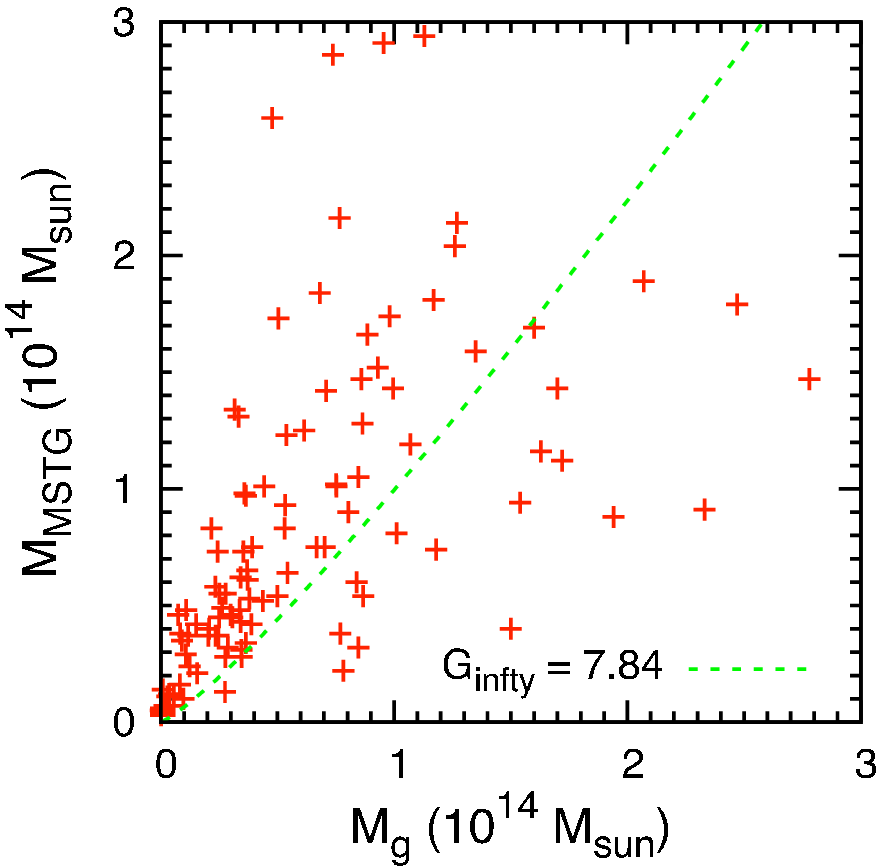}
\end{tabular}
\end{center}
\caption{A plot of the observed masses of the galaxy clusters in our sample listed in Table~\ref{clusterProperties}
versus the inferred Newtonian dynamic masses of the galaxy clusters and the best fit MSTG dynamic masses renormalized by
three different schemes across all galaxy clusters: (1) $M_{0} \propto M_{\rm gas}^{a}$, where $M_{\rm gas}$ is the total mass
of the cluster (black long dashed
curve), (2) $M_{0} \equiv \rm{constant}$ (blue dotted curve), and (3) $G_{\infty} \equiv \rm{constant}$ (green short
dashed curve). Red $+$-points are individual clusters from $\beta-$model data~\citep{rei01, rei02}. The
MSTG dynamic mass in these three schemes is
plotted in the remaining three graphs. Scheme (1) leads to the smallest sum of squares and is the overall best fit to
the data with $M_{0} \approx 58.8 M_{\rm gas}^{0.39} [10^{14} M_{\sun}]$. Scheme (2) is an acceptable fit of the bulk
properties with $M_{0} \approx 59.4 \times 10^{14} M_{\sun}$. Scheme (3) is a marginal fit of the bulk properties with
$G_{\infty} \approx 7.84 G_{0}$. Both dynamic masses, $M_{\rm MSTG}$ and $M_{\rm N}$ are computed for isotropic,
isothermal spheres cutoff at a spatial extent corresponding to a density of $10^{-28}\,\mbox{g/cm}^{3}$. 
Error bars are omitted for clarity -- the computed errors are provided in Table~\ref{clusterProperties} for
completeness.\label{dynamicMass}}
\end{figure*}
\clearpage

\setcounter{figure}{1}
\begin{figure*}
\begin{center}
\begin{tabular}{c}
\includegraphics[width=165mm]{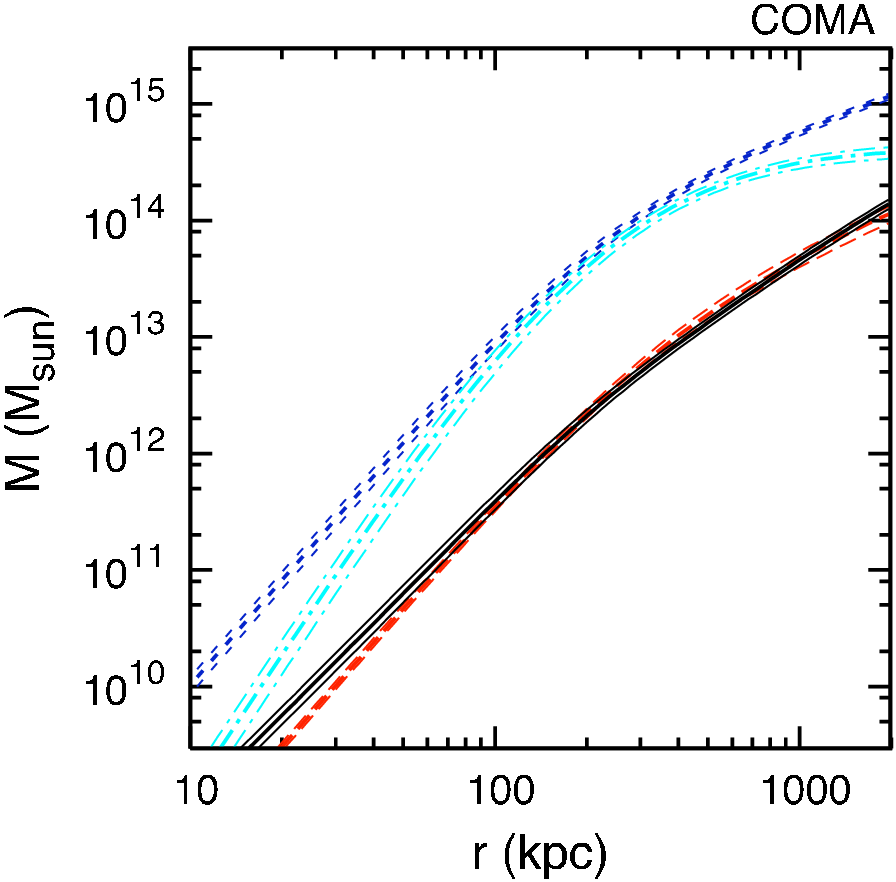}
\end{tabular}
\end{center}
\caption{Galaxy Cluster Mass Profiles: Plot of the radial mass profile for the 106 clusters in the sample of
Table~\ref{clusterProperties}. In all cases, the horizontal axis is the radius in kpc and the vertical axis is mass in
units of $10^{14} M_{\sun}$. The red long dashed curve is the ICM gas mass inferred from X-ray observations according to
the compilation of \citet{rei01, rei02}; the short dashed blue curve is the Newtonian dynamic mass; the dashed-dotted
cyan curve is the MOND dynamic mass; and the solid black curve is the MSTG dynamic mass. The Newtonian, MOND and MSTG
dynamic masses are calculated within the context of the $\beta-$model isothermal, isotropic sphere. The uncertainty in
$T$, $\beta$, $r_{c}$ and $H_{0}$ is shown for the COMA cluster as a representative case of a cluster with a well fit $\beta$-model; but error bands are omitted in the
subsequent thumbnail plots for clarity.\label{clusters}}
\end{figure*}
\clearpage

\setcounter{figure}{2}
\begin{figure*}
\begin{center}
\begin{tabular}{ccc}
\includegraphics[width=50mm]{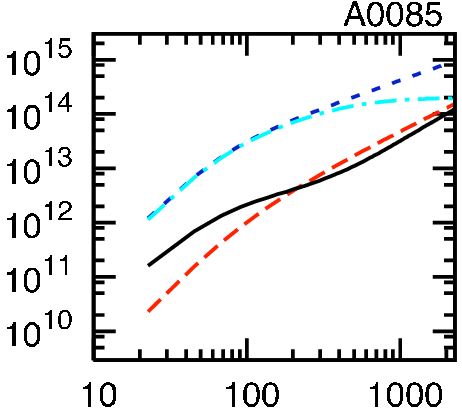} &
\includegraphics[width=50mm]{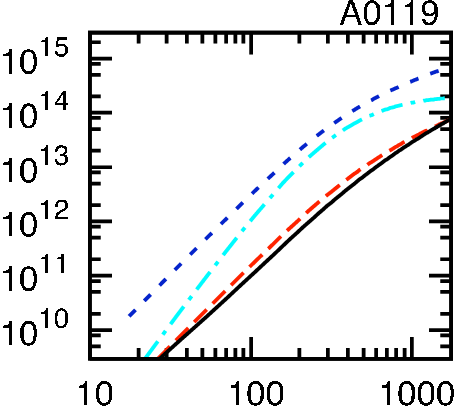} &
\includegraphics[width=50mm]{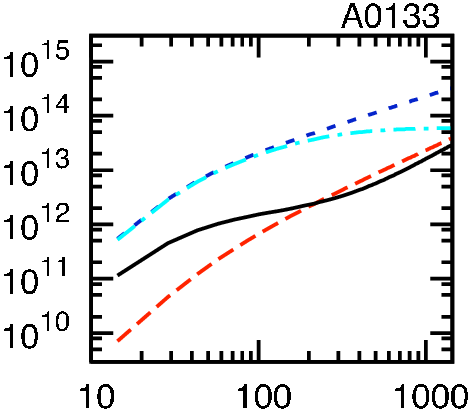} \\
\includegraphics[width=50mm]{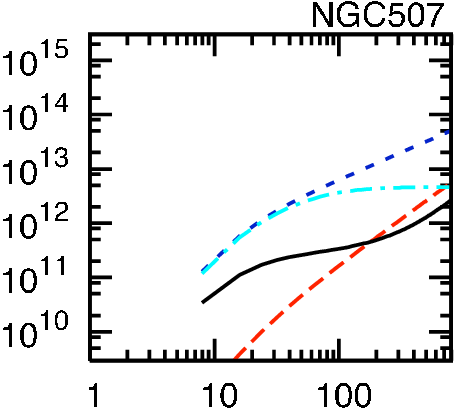} &
\includegraphics[width=50mm]{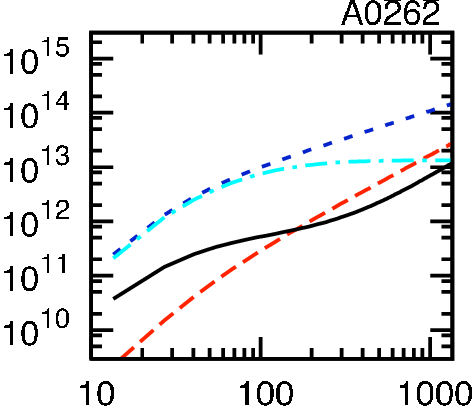} &
\includegraphics[width=50mm]{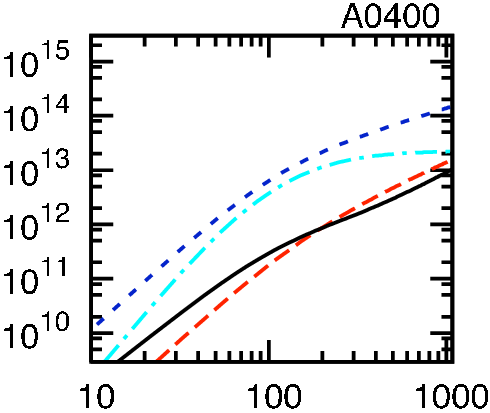} \\
\includegraphics[width=50mm]{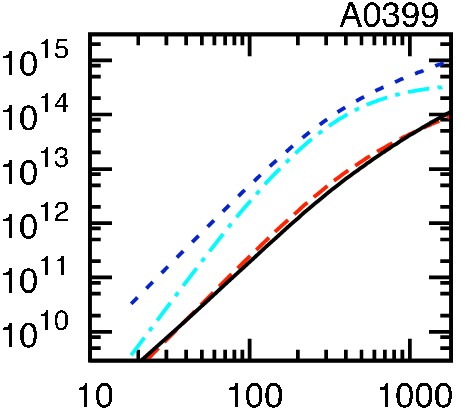} &
\includegraphics[width=50mm]{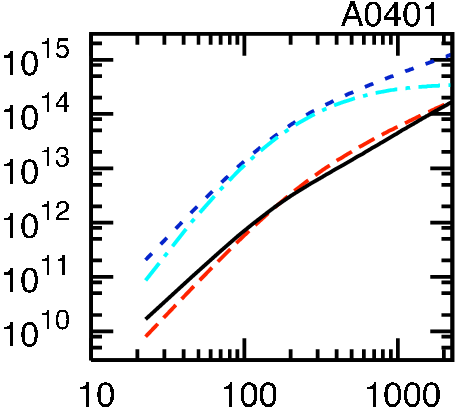} &
\includegraphics[width=50mm]{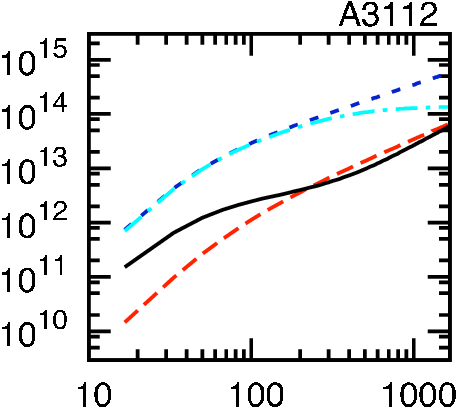} \\
\includegraphics[width=50mm]{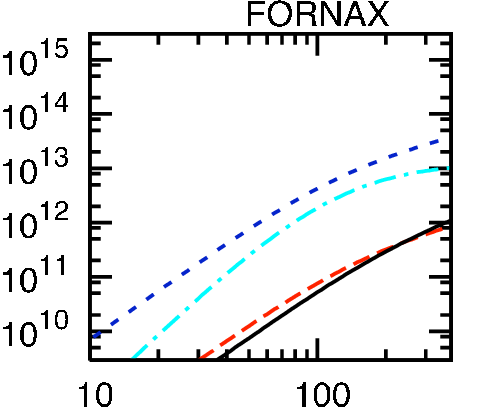} &
\includegraphics[width=50mm]{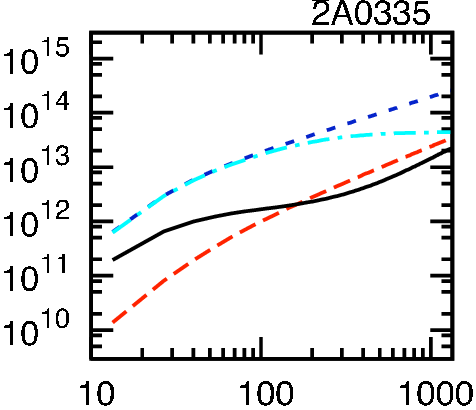} &
\includegraphics[width=50mm]{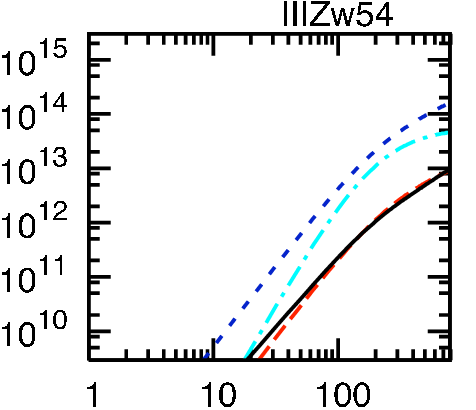} \\
\includegraphics[width=50mm]{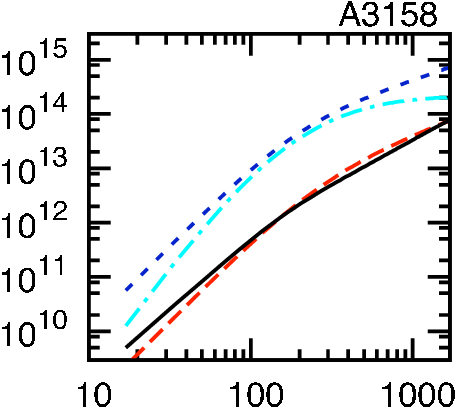} &
\includegraphics[width=50mm]{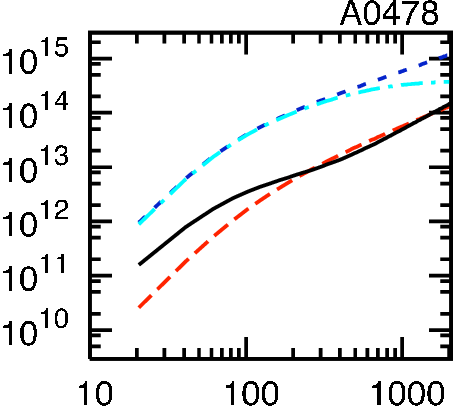} &
\includegraphics[width=50mm]{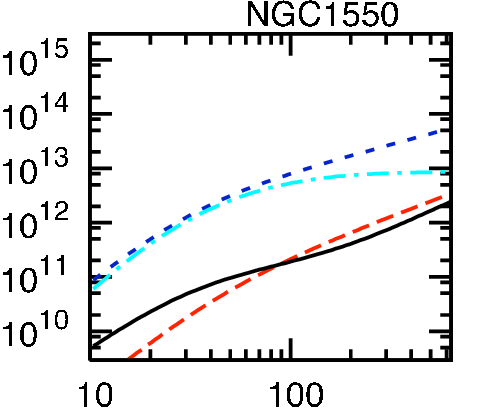}
\end{tabular}
\end{center}
\contcaption{Galaxy Cluster Mass Profiles}
\end{figure*}
\clearpage

\setcounter{figure}{2}
\begin{figure*}
\begin{center}
\begin{tabular}{ccc}
\includegraphics[width=50mm]{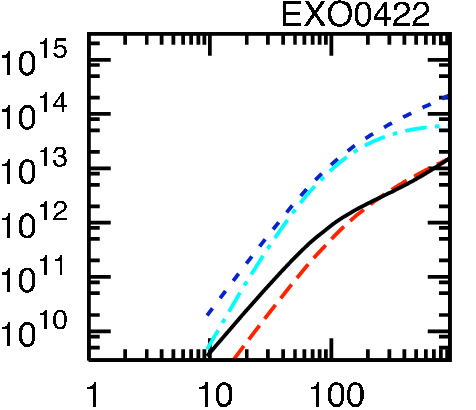} &
\includegraphics[width=50mm]{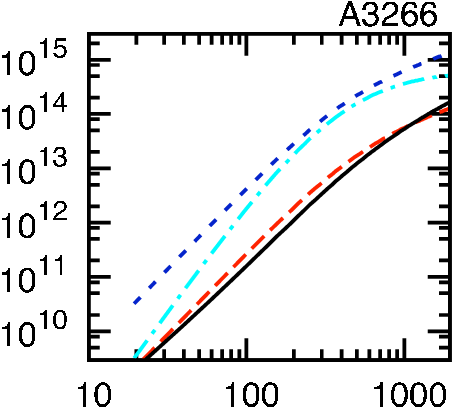} &
\includegraphics[width=50mm]{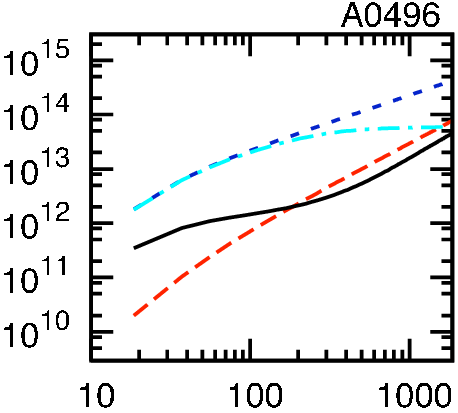} \\
\includegraphics[width=50mm]{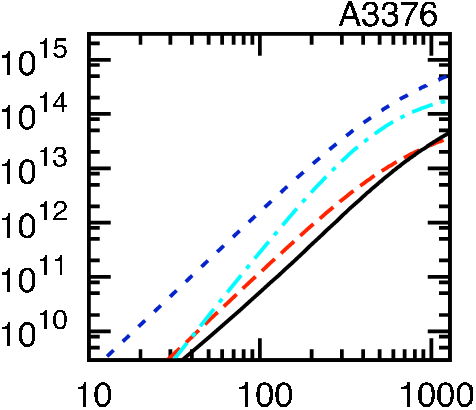} &
\includegraphics[width=50mm]{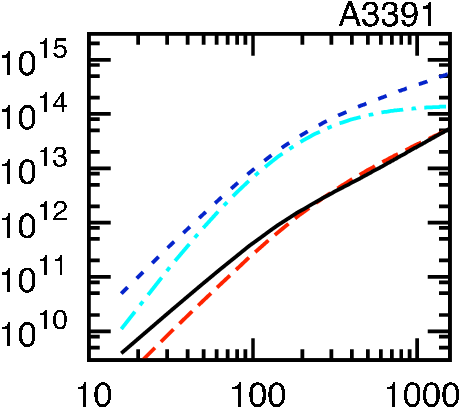} &
\includegraphics[width=50mm]{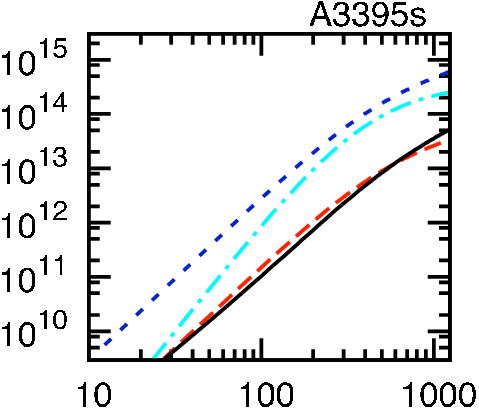} \\
\includegraphics[width=50mm]{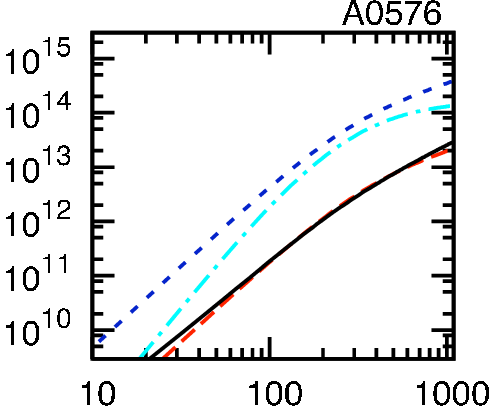} &
\includegraphics[width=50mm]{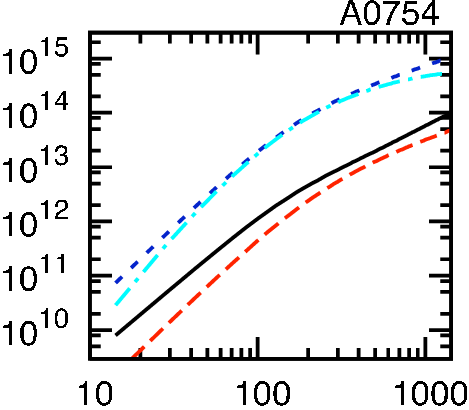} &
\includegraphics[width=50mm]{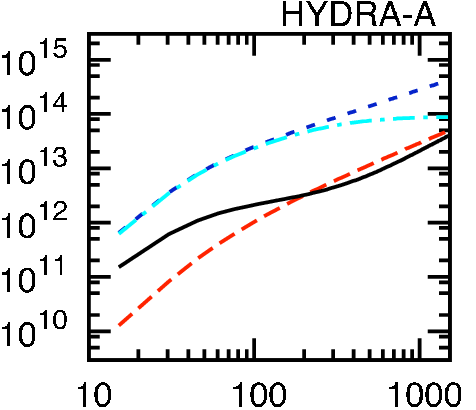} \\
\includegraphics[width=50mm]{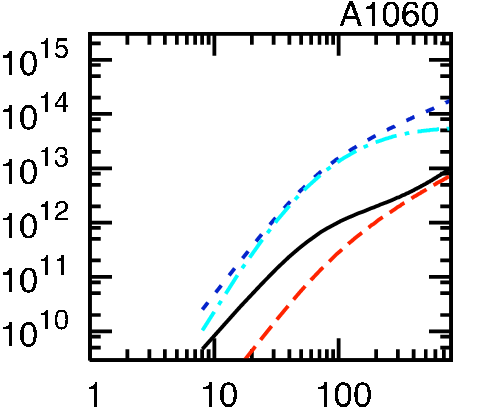} &
\includegraphics[width=50mm]{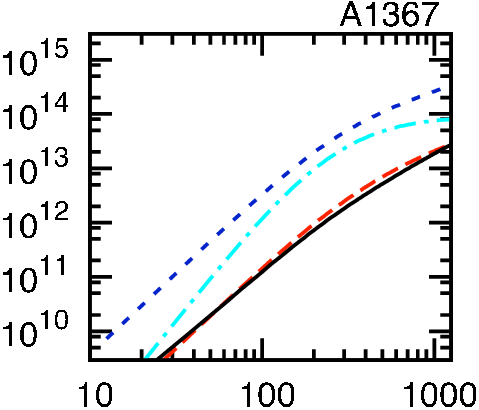} &
\includegraphics[width=50mm]{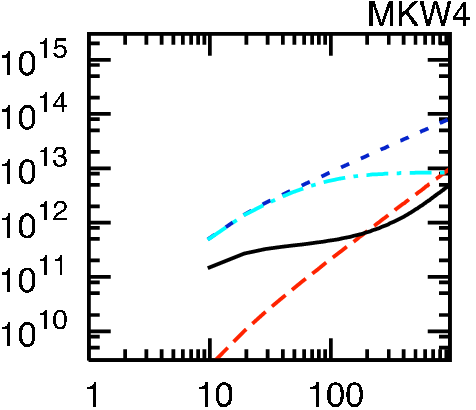} \\
\includegraphics[width=50mm]{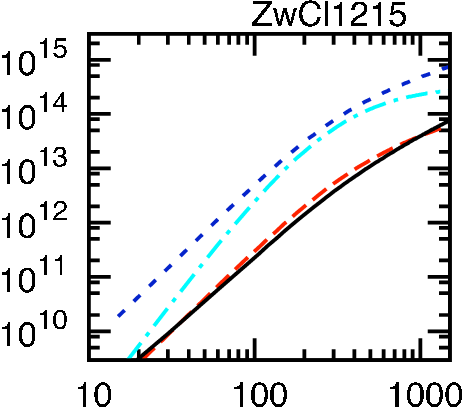} &
\includegraphics[width=50mm]{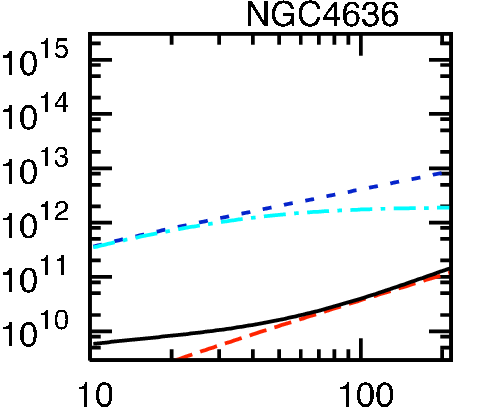} &
\includegraphics[width=50mm]{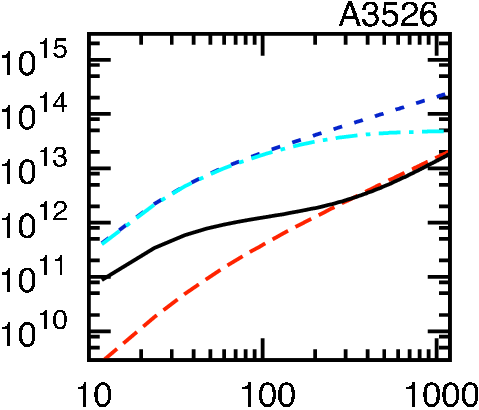}
\end{tabular}
\end{center}
\contcaption{Galaxy Cluster Mass Profiles}
\end{figure*}
\clearpage

\setcounter{figure}{2}
\begin{figure*}
\begin{center}
\begin{tabular}{ccc}
\includegraphics[width=50mm]{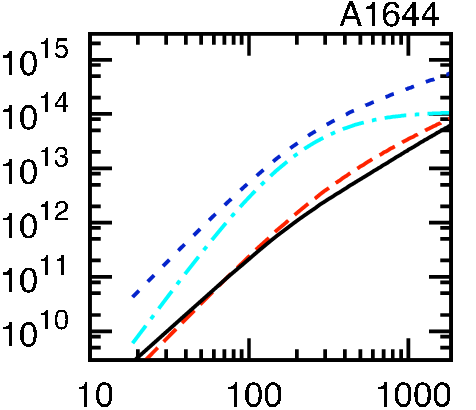} &
\includegraphics[width=50mm]{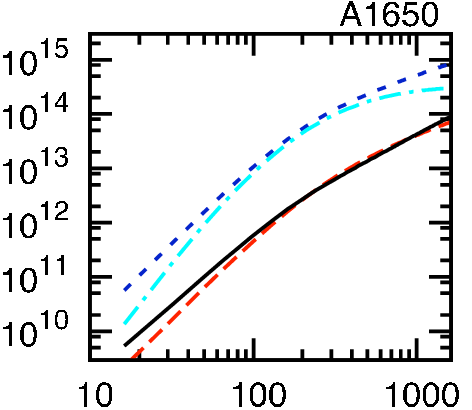} &
\includegraphics[width=50mm]{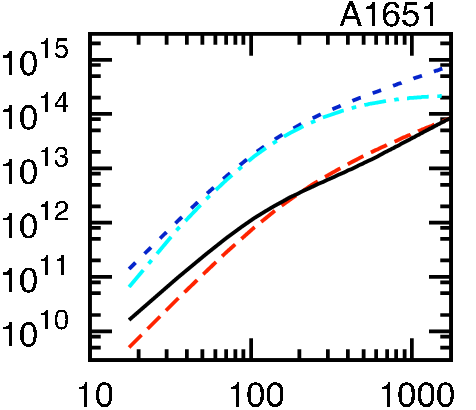} \\
\includegraphics[width=50mm]{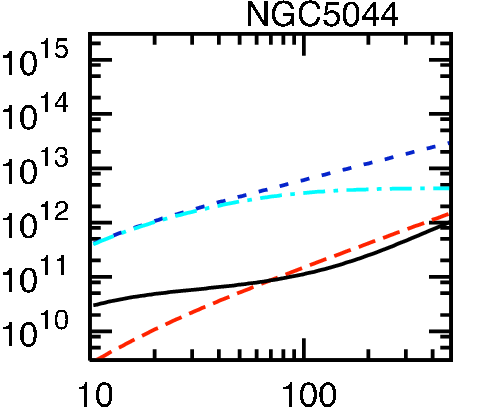} &
\includegraphics[width=50mm]{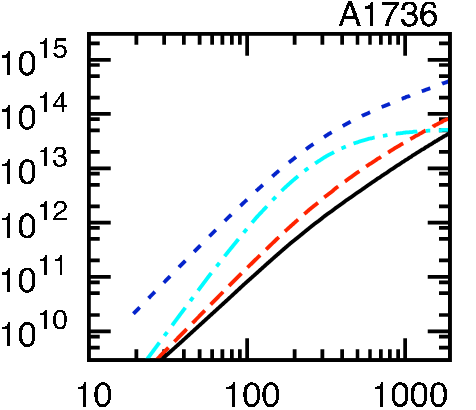} &
\includegraphics[width=50mm]{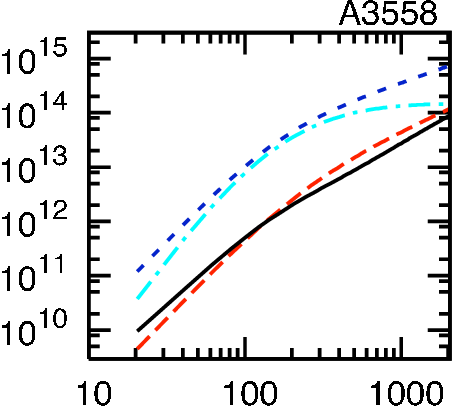} \\
\includegraphics[width=50mm]{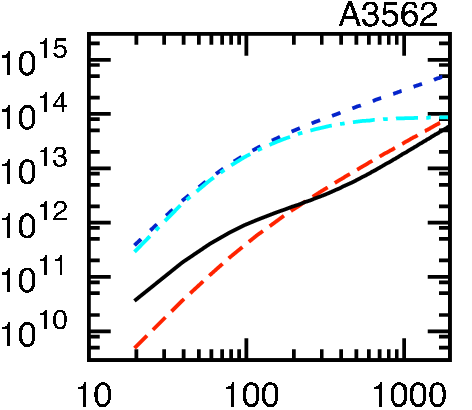} &
\includegraphics[width=50mm]{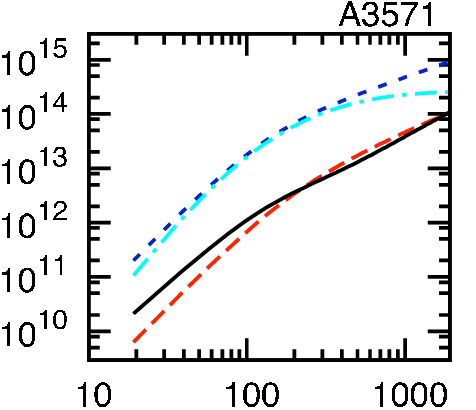} &
\includegraphics[width=50mm]{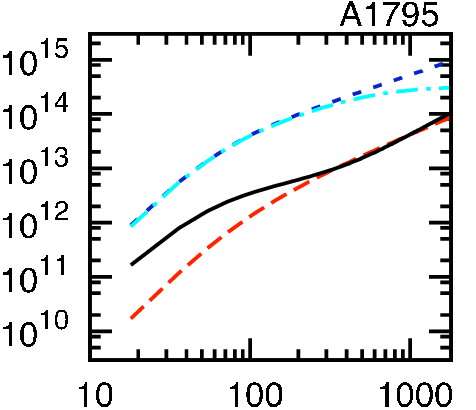} \\
\includegraphics[width=50mm]{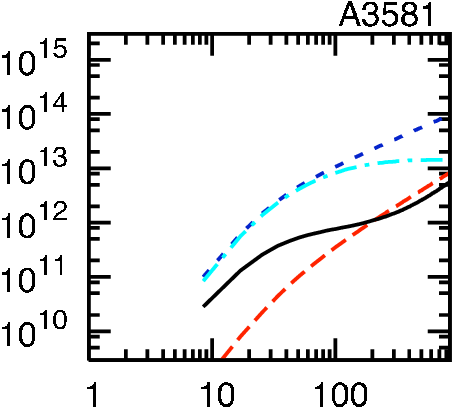} &
\includegraphics[width=50mm]{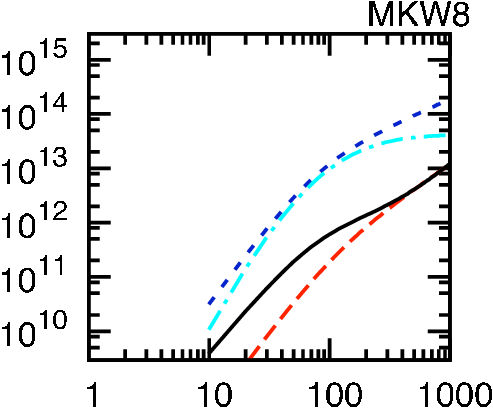} &
\includegraphics[width=50mm]{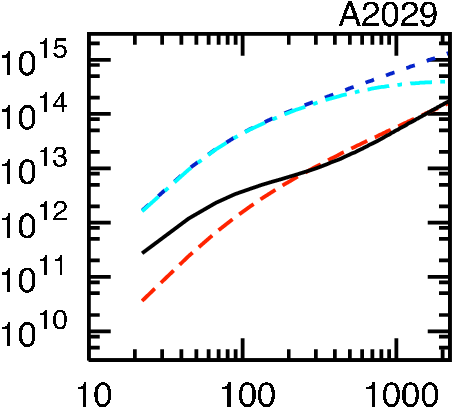} \\
\includegraphics[width=50mm]{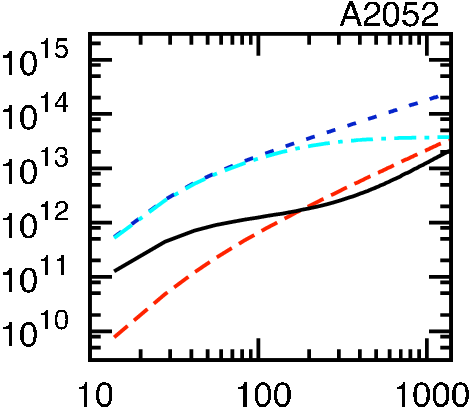} &
\includegraphics[width=50mm]{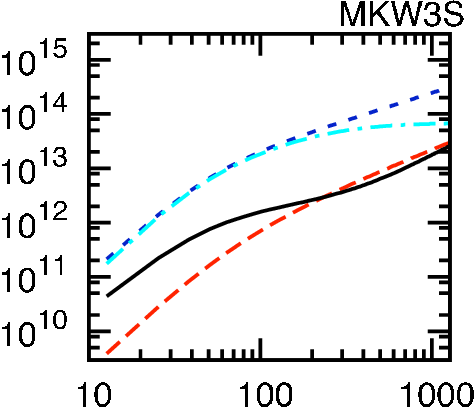} &
\includegraphics[width=50mm]{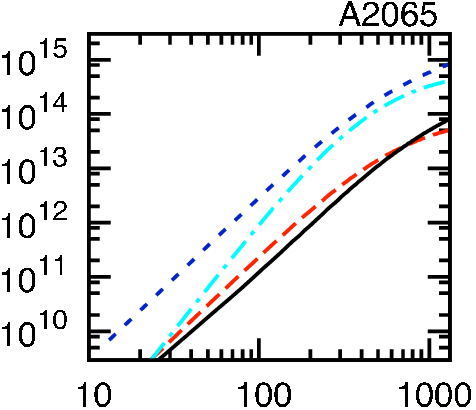}
\end{tabular}
\end{center}
\contcaption{Galaxy Cluster Mass Profiles}
\end{figure*}
\clearpage

\setcounter{figure}{2}
\begin{figure*}
\begin{center}
\begin{tabular}{ccc}
\includegraphics[width=50mm]{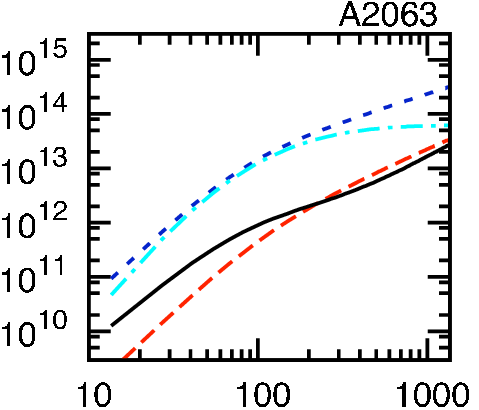} &
\includegraphics[width=50mm]{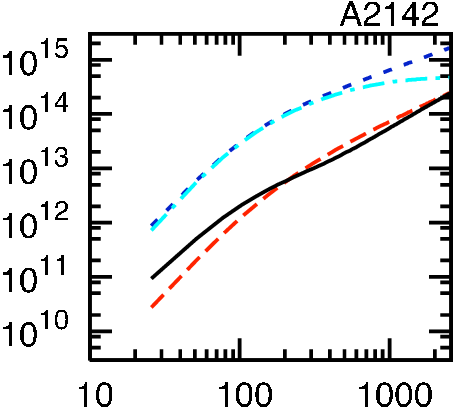} &
\includegraphics[width=50mm]{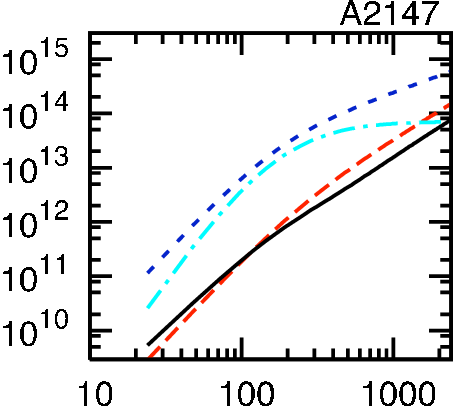} \\
\includegraphics[width=50mm]{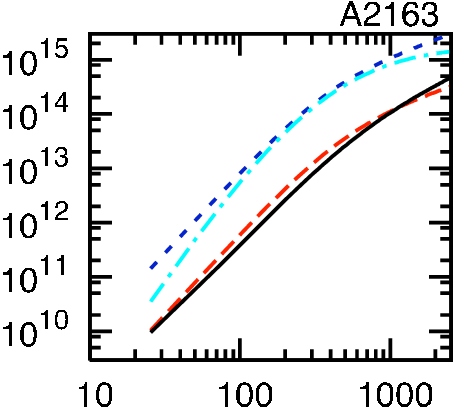} &
\includegraphics[width=50mm]{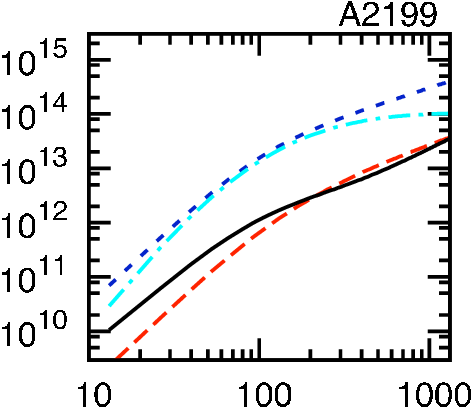} &
\includegraphics[width=50mm]{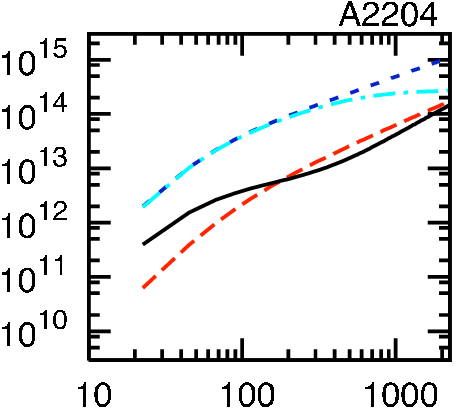} \\
\includegraphics[width=50mm]{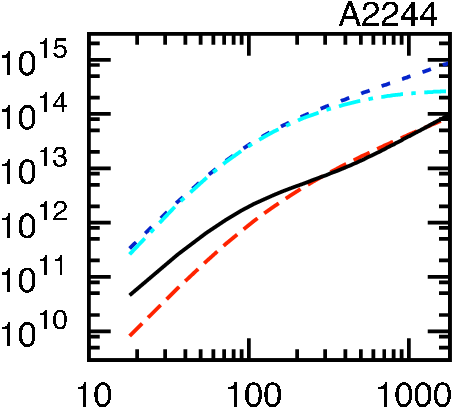} &
\includegraphics[width=50mm]{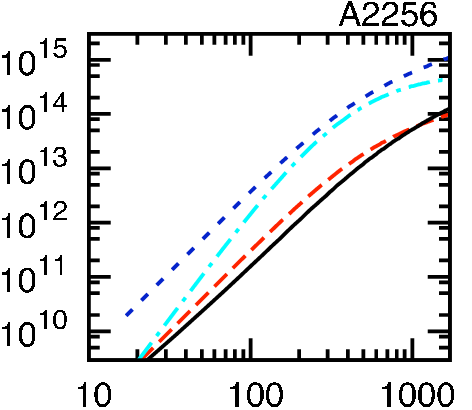} &
\includegraphics[width=50mm]{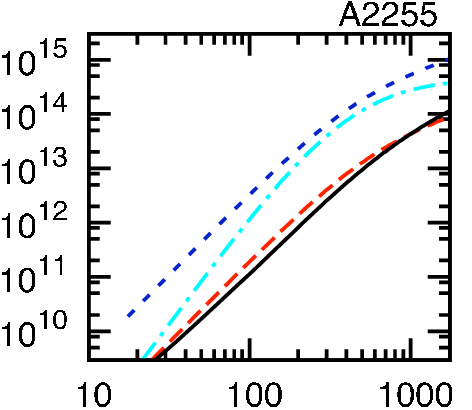} \\
\includegraphics[width=50mm]{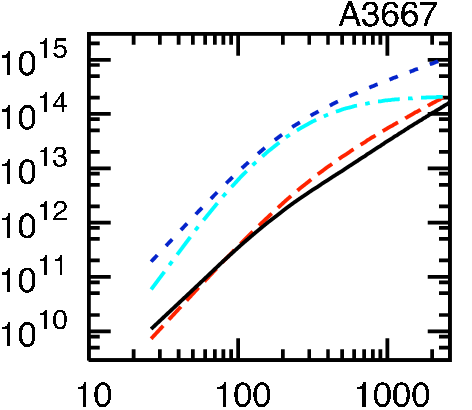} &
\includegraphics[width=50mm]{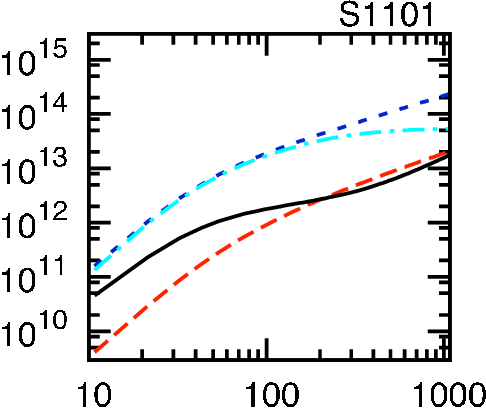} &
\includegraphics[width=50mm]{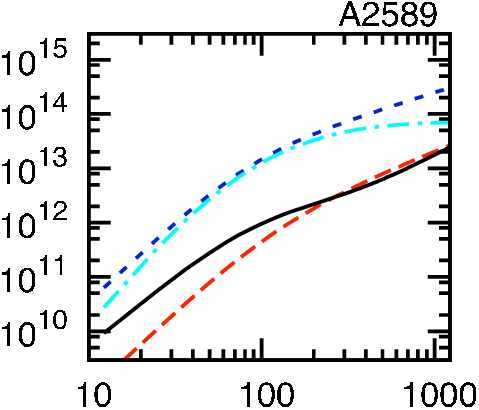} \\
\includegraphics[width=50mm]{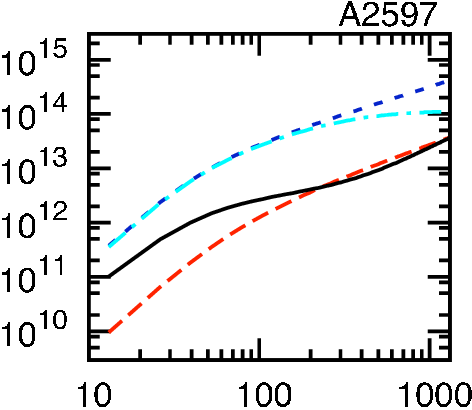} &
\includegraphics[width=50mm]{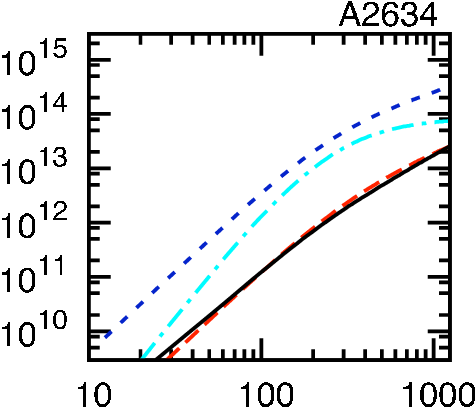} &
\includegraphics[width=50mm]{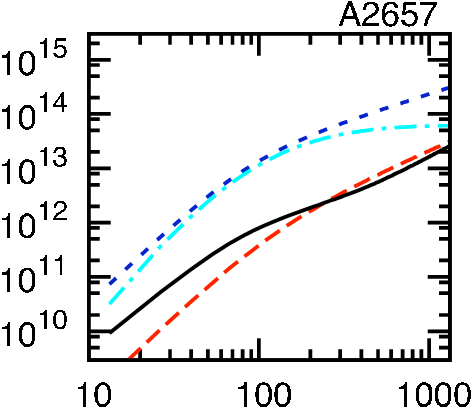}
\end{tabular}
\end{center}
\contcaption{Galaxy Cluster Mass Profiles}
\end{figure*}
\clearpage

\setcounter{figure}{2}
\begin{figure*}
\begin{center}
\begin{tabular}{ccc}
\includegraphics[width=50mm]{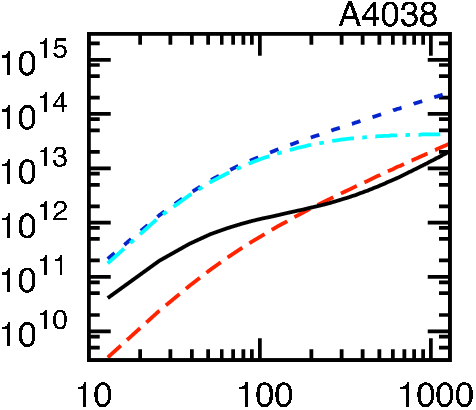} &
\includegraphics[width=50mm]{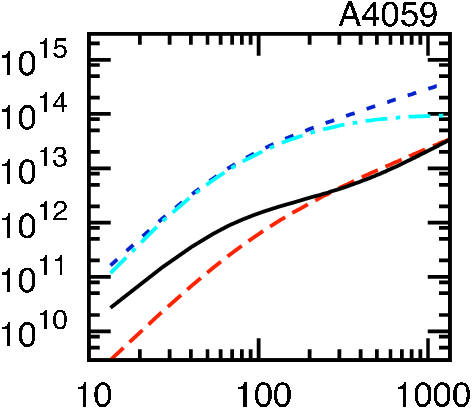} &
\includegraphics[width=50mm]{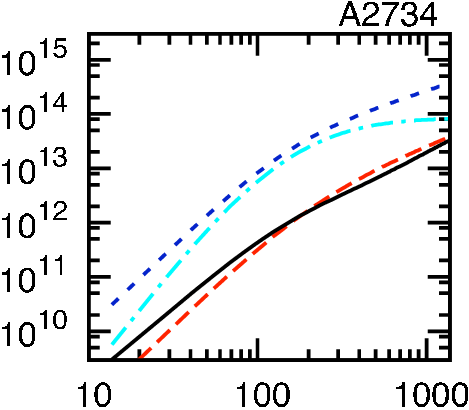} \\
\includegraphics[width=50mm]{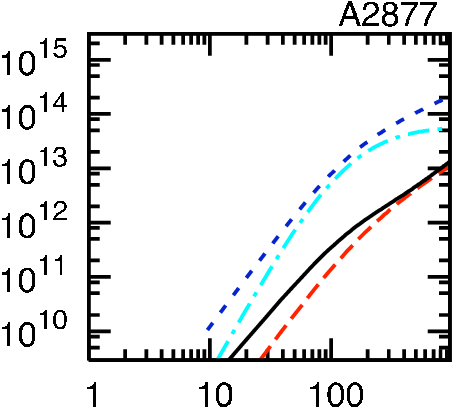} &
\includegraphics[width=50mm]{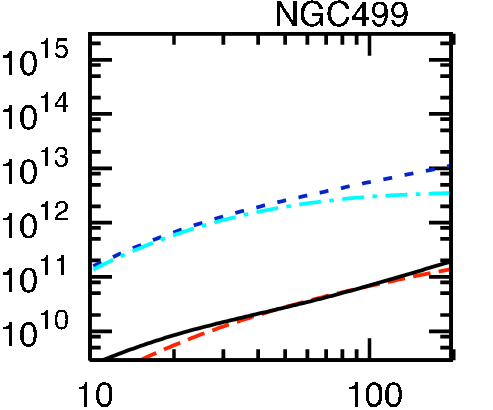} &
\includegraphics[width=50mm]{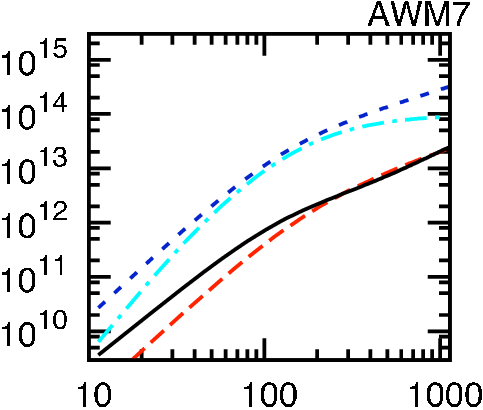} \\
\includegraphics[width=50mm]{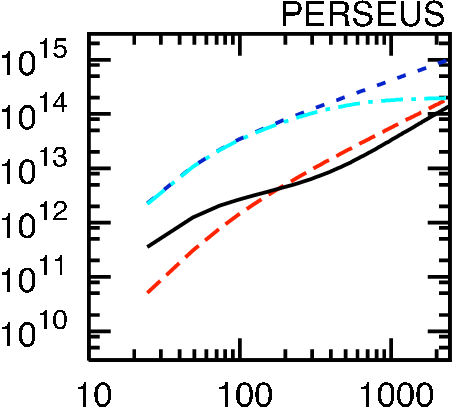} &
\includegraphics[width=50mm]{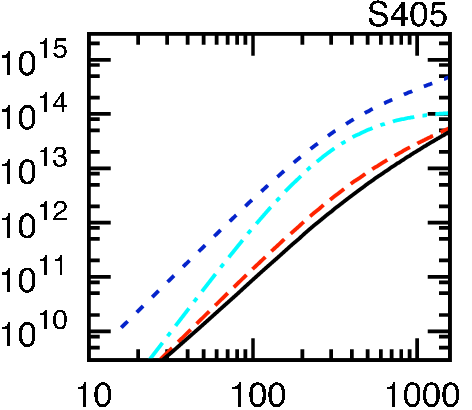} &
\includegraphics[width=50mm]{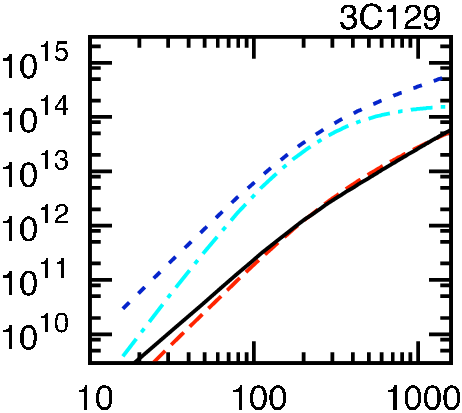} \\
\includegraphics[width=50mm]{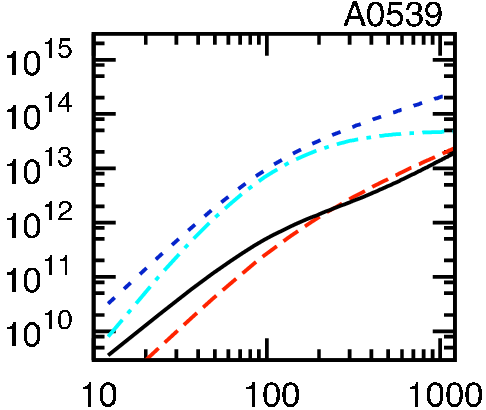} &
\includegraphics[width=50mm]{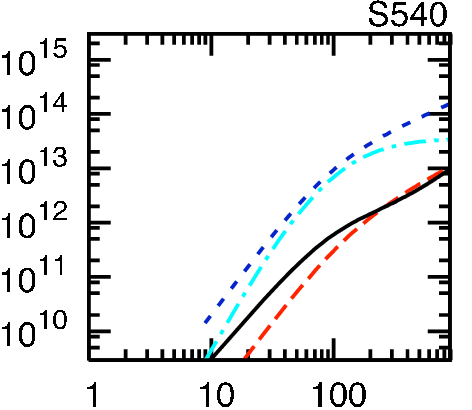} &
\includegraphics[width=50mm]{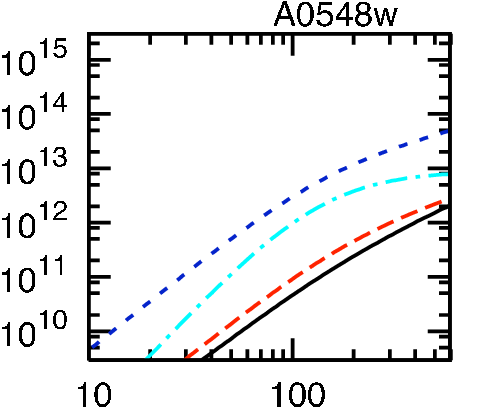} \\
\includegraphics[width=50mm]{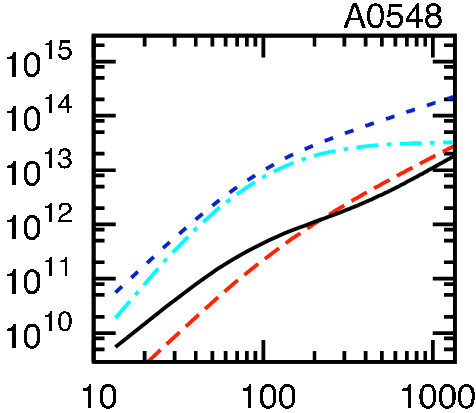} &
\includegraphics[width=50mm]{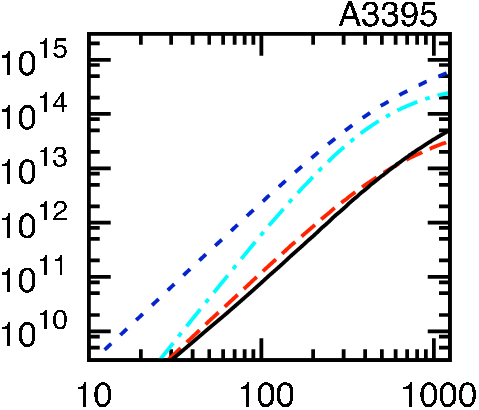} &
\includegraphics[width=50mm]{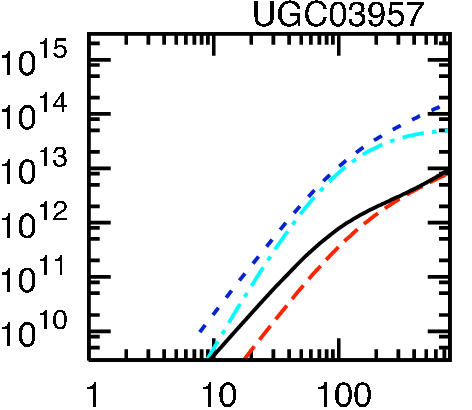}
\end{tabular}
\end{center}
\contcaption{Galaxy Cluster Mass Profiles}
\end{figure*}
\clearpage

\setcounter{figure}{2}
\begin{figure*}
\begin{center}
\begin{tabular}{ccc}
\includegraphics[width=50mm]{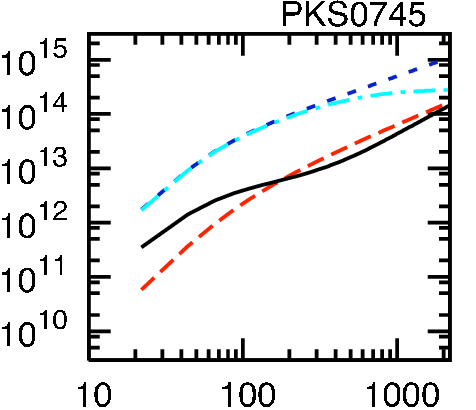} &
\includegraphics[width=50mm]{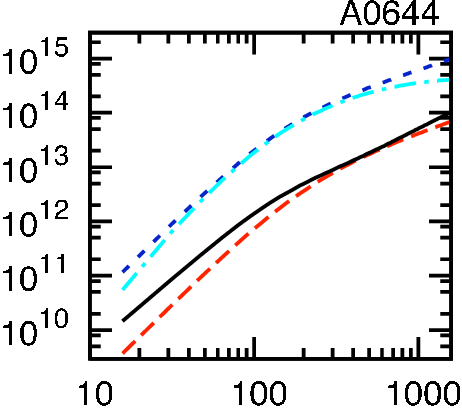} &
\includegraphics[width=50mm]{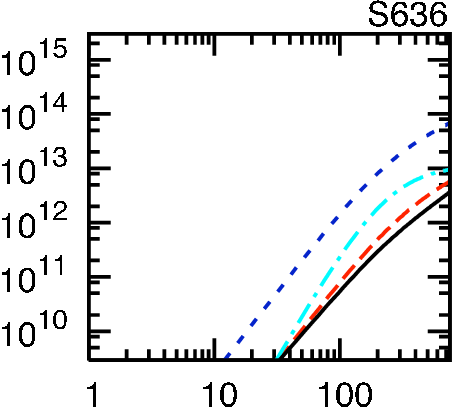} \\
\includegraphics[width=50mm]{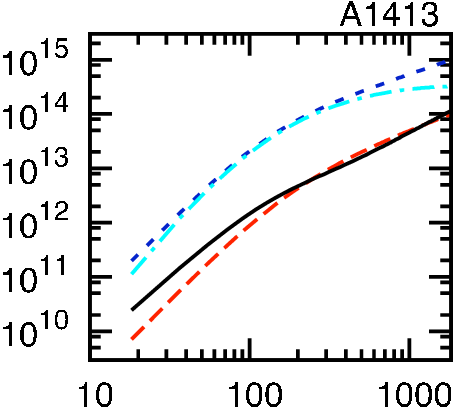} &
\includegraphics[width=50mm]{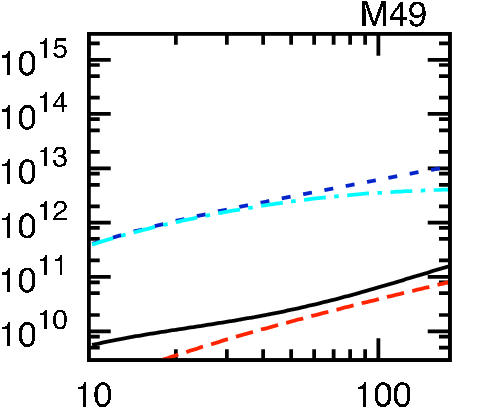} &
\includegraphics[width=50mm]{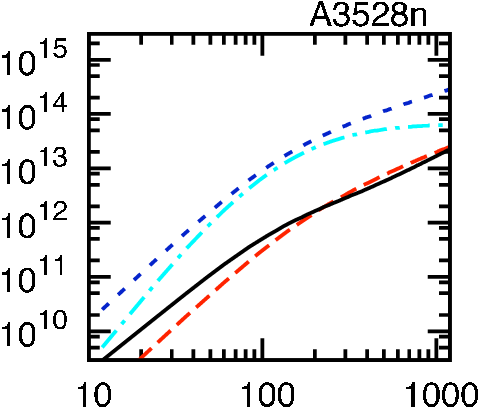} \\
\includegraphics[width=50mm]{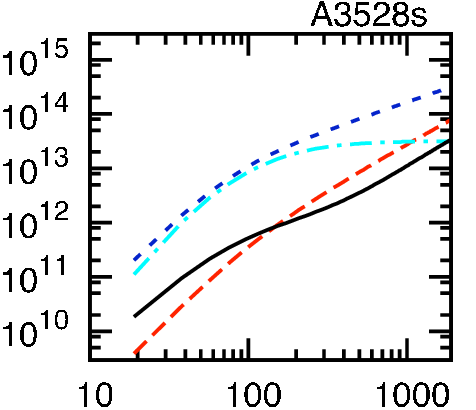} &
\includegraphics[width=50mm]{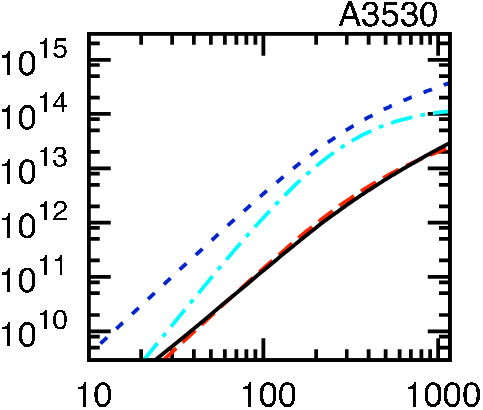} &
\includegraphics[width=50mm]{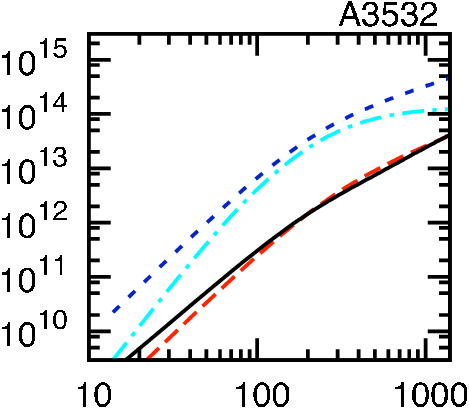} \\
\includegraphics[width=50mm]{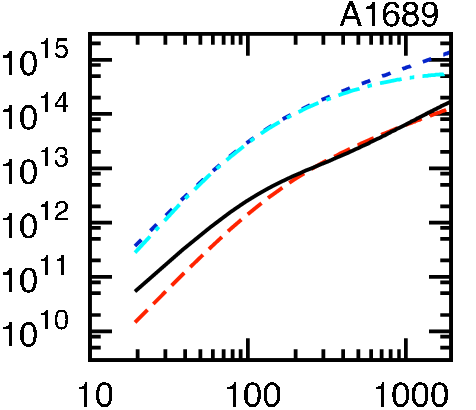} &
\includegraphics[width=50mm]{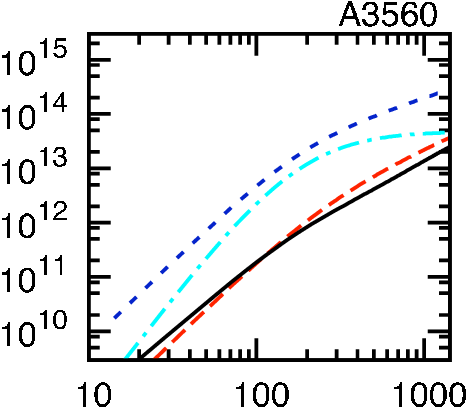} &
\includegraphics[width=50mm]{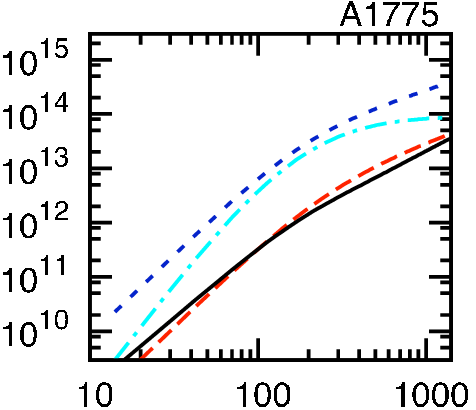} \\
\includegraphics[width=50mm]{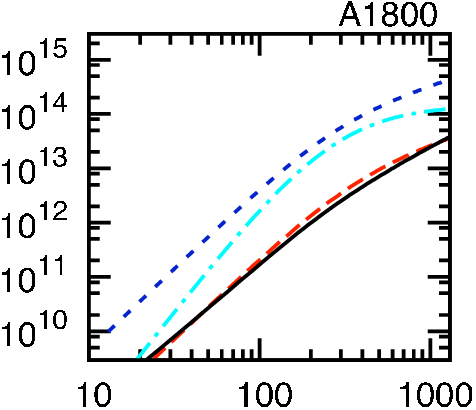} &
\includegraphics[width=50mm]{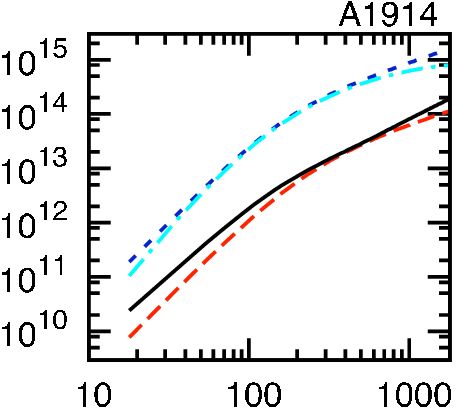} &
\includegraphics[width=50mm]{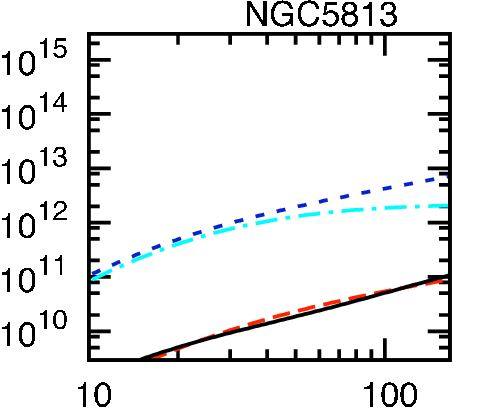}
\end{tabular}
\end{center}
\contcaption{Galaxy Cluster Mass Profiles}
\end{figure*}
\clearpage

\setcounter{figure}{2}
\begin{figure*}
\begin{center}
\begin{tabular}{ccc}
\includegraphics[width=50mm]{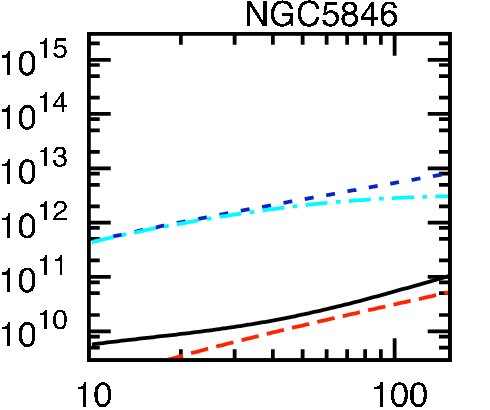} &
\includegraphics[width=50mm]{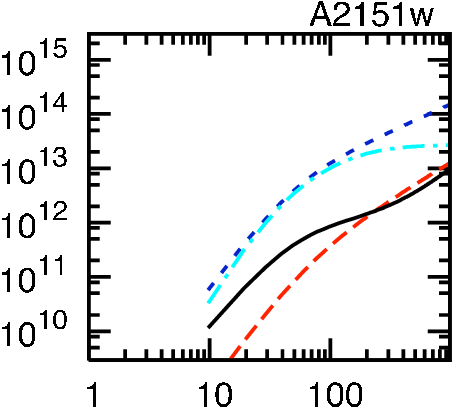} &
\includegraphics[width=50mm]{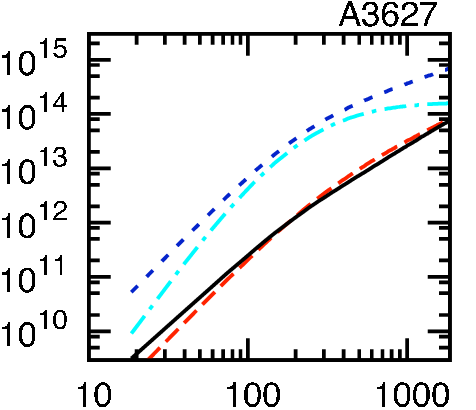} \\
\includegraphics[width=50mm]{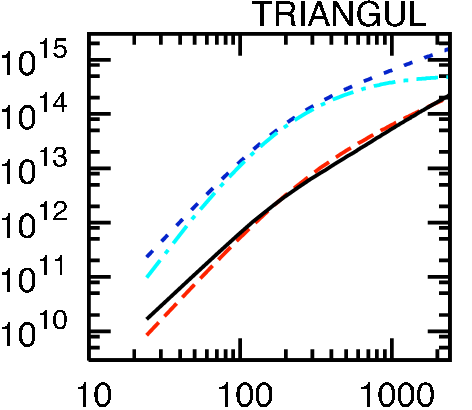} &
\includegraphics[width=50mm]{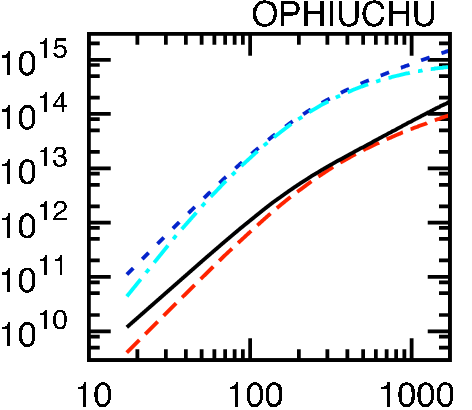} &
\includegraphics[width=50mm]{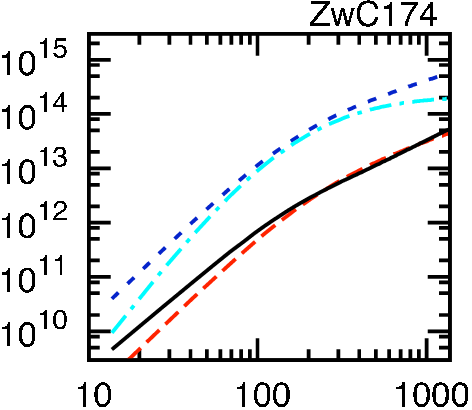} \\
\includegraphics[width=50mm]{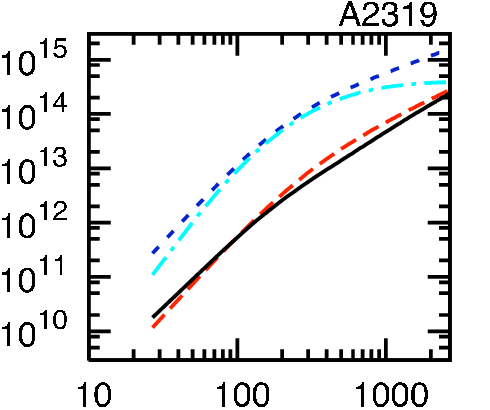} &
\includegraphics[width=50mm]{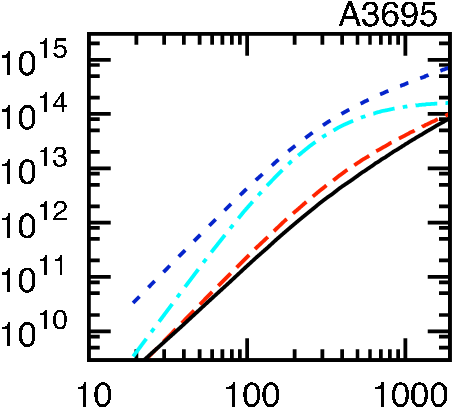} &
\includegraphics[width=50mm]{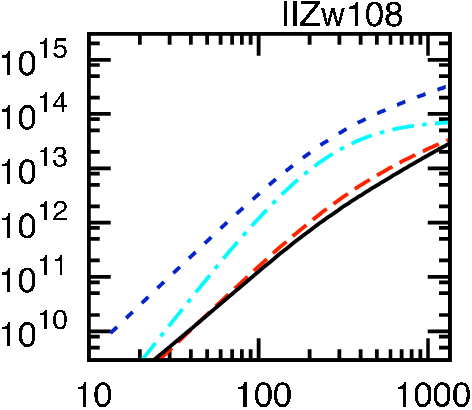} \\
\includegraphics[width=50mm]{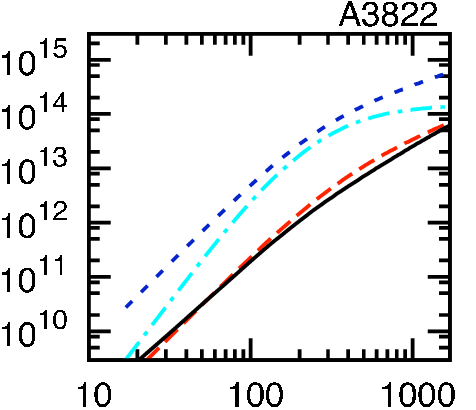} &
\includegraphics[width=50mm]{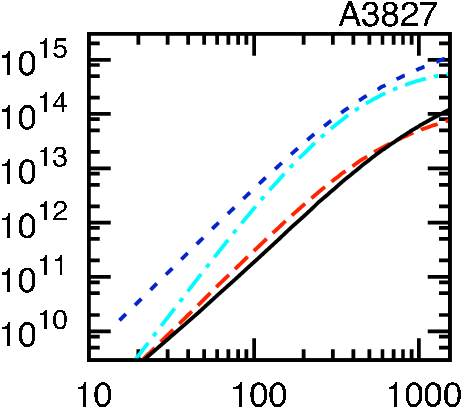} &
\includegraphics[width=50mm]{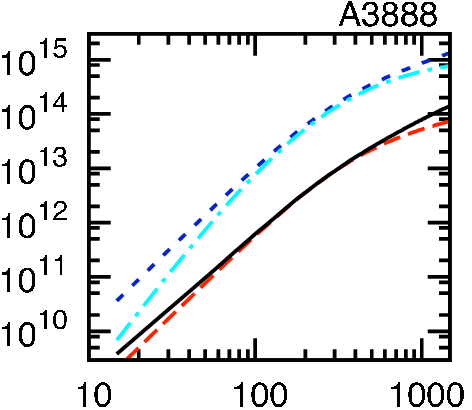} \\
\includegraphics[width=50mm]{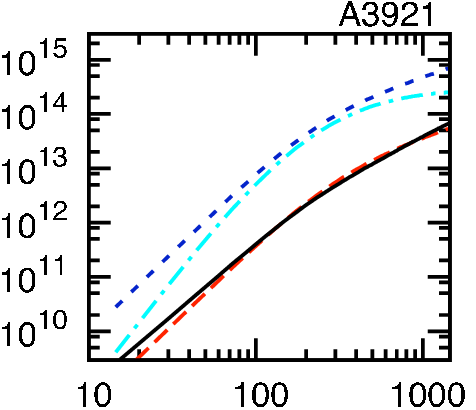} &
\includegraphics[width=50mm]{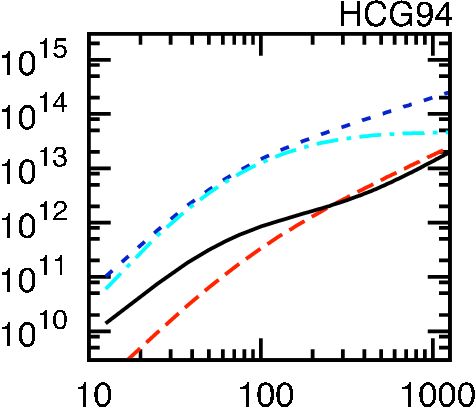} &
\includegraphics[width=50mm]{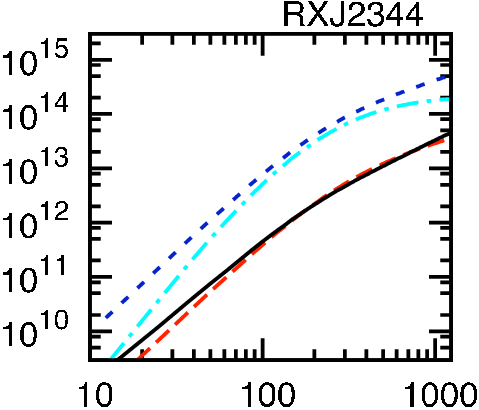}
\end{tabular}
\end{center}
\contcaption{Galaxy Cluster Mass Profiles}
\end{figure*}
\clearpage

\setcounter{table}{0}
\begin{table*}
\caption{\sc X-ray Cluster Properties of the {\em Complete Sample}:}
\label{clusterProperties}
\begin{tabular}{lcccrrcccc}
\hline\hline
\colhead{Cluster} & \colhead{$T$} & \colhead{$\rho_{0}$} & \colhead{$\beta$} & \colhead{$r_{c}$} & \colhead{$r_{\rm out}$} & \colhead{$M_{\rm gas}$} & \colhead{$M_{\rm N}$} & \colhead{$M_{\rm MSTG}$} & \colhead{$M_{\rm MOND}$}\\
& \colhead{[keV]} & \colhead{[$10^{-25}\,\mbox{g/cm}^{3}$]} && \colhead{[kpc]} & \colhead{[kpc]} & \colhead{[$10^{14} M_{\sun}$]} & \colhead{[$10^{14} M_{\sun}$]} & \colhead{[$10^{14} M_{\sun}$]} & \colhead{[$10^{14} M_{\sun}$]}\\
\colhead{\footnotesize (1)}&\colhead{\footnotesize (2)}&\colhead{\footnotesize (3)}&\colhead{\footnotesize (4)}&\colhead{\footnotesize (5)}&\colhead{\footnotesize (6)}&\colhead{\footnotesize (7)}&\colhead{\footnotesize (8)}&\colhead{\footnotesize (9)}&\colhead{\footnotesize (10)}\\
\hline
A0085&$6.90^{+0.40}_{-0.40}$&0.34&$0.532^{+0.004}_{-0.004}$&$58.5^{+3.3}_{-3.9}$&$2,241_{-162}^{+139}$&$1.48_{-0.16}^{+0.14}$&$9.02_{-0.53}^{+0.53}$&$1.15_{-0.09}^{+0.08}$&$1.83_{-0.21}^{+0.21}$\\
A0119&$5.60^{+0.30}_{-0.30}$&0.03&$0.675^{+0.026}_{-0.023}$&$352.8^{+24.7}_{-27.0}$&$1,728_{-173}^{+156}$&$0.73_{-0.10}^{+0.10}$&$6.88_{-0.50}^{+0.53}$&$0.73_{-0.07}^{+0.07}$&$1.76_{-0.25}^{+0.26}$\\
A0133&$3.80^{+2.00}_{-0.90}$&0.42&$0.530^{+0.004}_{-0.004}$&$31.7^{+1.9}_{-2.3}$&$1,417_{-109}^{+96}$&$0.37_{-0.04}^{+0.04}$&$3.13_{-0.74}^{+1.65}$&$0.28_{-0.07}^{+0.15}$&$0.55_{-0.26}^{+0.57}$\\
NGC507&$1.26^{+0.07}_{-0.07}$&0.23&$0.444^{+0.005}_{-0.005}$&$13.4^{+0.9}_{-1.0}$&$783_{-70}^{+64}$&$0.05_{-0.01}^{+0.01}$&$0.48_{-0.03}^{+0.03}$&$0.02_{-0.00}^{+0.00}$&$0.04_{-0.00}^{+0.00}$\\
A0262&$2.15^{+0.06}_{-0.06}$&0.16&$0.443^{+0.018}_{-0.017}$&$29.6^{+8.5}_{-7.2}$&$1,334_{-386}^{+432}$&$0.26_{-0.10}^{+0.11}$&$1.39_{-0.09}^{+0.09}$&$0.11_{-0.02}^{+0.02}$&$0.13_{-0.02}^{+0.02}$\\
A0400&$2.31^{+0.14}_{-0.14}$&0.04&$0.534^{+0.014}_{-0.013}$&$108.5^{+7.8}_{-8.8}$&$1,062_{-108}^{+97}$&$0.15_{-0.02}^{+0.02}$&$1.42_{-0.10}^{+0.10}$&$0.10_{-0.01}^{+0.01}$&$0.20_{-0.03}^{+0.03}$\\
A0399&$7.00^{+0.40}_{-0.40}$&0.04&$0.713^{+0.137}_{-0.095}$&$316.9^{+93.9}_{-72.7}$&$1,791_{-744}^{+683}$&$0.90_{-0.48}^{+0.53}$&$9.51_{-1.88}^{+2.64}$&$1.07_{-0.33}^{+0.41}$&$3.07_{-1.15}^{+1.62}$\\
A0401&$8.00^{+0.40}_{-0.40}$&0.11&$0.613^{+0.010}_{-0.010}$&$173.2^{+10.7}_{-12.0}$&$2,236_{-182}^{+167}$&$1.65_{-0.22}^{+0.20}$&$11.96_{-0.66}^{+0.66}$&$1.58_{-0.13}^{+0.12}$&$3.18_{-0.34}^{+0.34}$\\
A3112&$5.30^{+0.70}_{-1.00}$&0.54&$0.576^{+0.006}_{-0.006}$&$43.0^{+2.8}_{-3.2}$&$1,644_{-138}^{+124}$&$0.64_{-0.09}^{+0.08}$&$5.50_{-1.04}^{+0.73}$&$0.56_{-0.11}^{+0.08}$&$1.26_{-0.46}^{+0.33}$\\
FORNAX&$1.20^{+0.04}_{-0.04}$&0.02&$0.804^{+0.098}_{-0.084}$&$122.5^{+13.0}_{-12.6}$&$387_{-74}^{+67}$&$.009_{-.003}^{+.004}$&$.373_{-.057}^{+.066}$&$.011_{-.003}^{+.003}$&$.102_{-.030}^{+.035}$\\
2A0335&$3.01^{+0.07}_{-0.07}$&1.07&$0.575^{+0.004}_{-0.003}$&$23.2^{+1.2}_{-1.5}$&$1,322_{-92}^{+74}$&$0.33_{-0.04}^{+0.03}$&$2.51_{-0.06}^{+0.06}$&$0.21_{-0.01}^{+0.01}$&$0.41_{-0.02}^{+0.02}$\\
IIIZw54&$2.16^{+0.35}_{-0.30}$&0.04&$0.887^{+0.320}_{-0.151}$&$203.5^{+87.7}_{-52.7}$&$780_{-461}^{+389}$&$0.09_{-0.09}^{+0.09}$&$1.54_{-0.43}^{+0.82}$&$0.09_{-0.09}^{+0.09}$&$0.43_{-0.23}^{+0.44}$\\
A3158&$5.77^{+0.10}_{-0.05}$&0.08&$0.661^{+0.025}_{-0.022}$&$189.4^{+16.2}_{-17.1}$&$1,672_{-206}^{+189}$&$0.73_{-0.14}^{+0.13}$&$6.91_{-0.33}^{+0.39}$&$0.73_{-0.07}^{+0.07}$&$1.89_{-0.17}^{+0.20}$\\
A0478&$8.40^{+0.80}_{-1.40}$&0.5&$0.613^{+0.004}_{-0.004}$&$69.0^{+3.2}_{-4.1}$&$2,029_{-130}^{+105}$&$1.30_{-0.15}^{+0.12}$&$11.45_{-1.91}^{+1.10}$&$1.42_{-0.25}^{+0.15}$&$3.49_{-1.11}^{+0.64}$\\
NGC1550&$1.43^{+0.04}_{-0.03}$&0.15&$0.554^{+0.049}_{-0.037}$&$31.7^{+10.6}_{-7.9}$&$632_{-231}^{+247}$&$.034_{-.018}^{+.020}$&$.548_{-.053}^{+.070}$&$.024_{-.007}^{+.008}$&$.086_{-.016}^{+.022}$\\
EXO0422&$2.90^{+0.90}_{-0.60}$&0.13&$0.722^{+0.104}_{-0.071}$&$100.0^{+28.5}_{-21.9}$&$934_{-367}^{+338}$&$0.15_{-0.09}^{+0.09}$&$2.12_{-0.53}^{+0.79}$&$0.14_{-0.05}^{+0.07}$&$0.57_{-0.27}^{+0.41}$\\
A3266&$8.00^{+0.50}_{-0.50}$&0.05&$0.796^{+0.020}_{-0.019}$&$397.2^{+22.4}_{-26.4}$&$1,915_{-150}^{+132}$&$1.22_{-0.16}^{+0.13}$&$12.82_{-0.91}^{+0.92}$&$1.56_{-0.14}^{+0.13}$&$4.79_{-0.63}^{+0.64}$\\
A0496&$4.13^{+0.08}_{-0.08}$&0.65&$0.484^{+0.003}_{-0.003}$&$21.1^{+1.1}_{-1.4}$&$1,830_{-130}^{+111}$&$0.74_{-0.07}^{+0.06}$&$4.01_{-0.09}^{+0.09}$&$0.43_{-0.02}^{+0.02}$&$0.55_{-0.02}^{+0.02}$\\
A3376&$4.00^{+0.40}_{-0.40}$&0.02&$1.054^{+0.101}_{-0.083}$&$531.7^{+53.5}_{-51.8}$&$1,264_{-183}^{+169}$&$0.35_{-0.35}^{+0.35}$&$4.97_{-0.76}^{+0.85}$&$0.43_{-0.43}^{+0.43}$&$1.67_{-0.48}^{+0.54}$\\
A3391&$5.40^{+0.60}_{-0.60}$&0.05&$0.579^{+0.026}_{-0.024}$&$164.8^{+18.3}_{-18.1}$&$1,558_{-234}^{+227}$&$0.51_{-0.11}^{+0.11}$&$5.28_{-0.66}^{+0.68}$&$0.51_{-0.08}^{+0.08}$&$1.29_{-0.31}^{+0.32}$\\
A3395s&$5.00^{+0.30}_{-0.30}$&0.03&$0.964^{+0.275}_{-0.167}$&$425.4^{+123.1}_{-86.5}$&$1,223_{-501}^{+442}$&$0.32_{-0.32}^{+0.32}$&$5.77_{-1.50}^{+2.37}$&$0.49_{-0.49}^{+0.49}$&$2.34_{-1.12}^{+1.76}$\\
A0576&$4.02^{+0.07}_{-0.07}$&0.03&$0.825^{+0.432}_{-0.185}$&$277.5^{+156.1}_{-89.4}$&$1,076_{-903}^{+703}$&$0.21_{-0.19}^{+0.41}$&$3.67_{-1.19}^{+2.72}$&$0.28_{-0.15}^{+0.32}$&$1.26_{-0.77}^{+1.76}$\\
A0754&$9.50^{+0.70}_{-0.40}$&0.09&$0.698^{+0.027}_{-0.024}$&$168.3^{+13.9}_{-14.7}$&$1,402_{-170}^{+156}$&$0.46_{-0.09}^{+0.08}$&$10.05_{-0.65}^{+0.92}$&$0.94_{-0.10}^{+0.12}$&$5.10_{-0.57}^{+0.82}$\\
HYDRA-A&$4.30^{+0.40}_{-0.40}$&0.63&$0.573^{+0.003}_{-0.003}$&$35.2^{+1.6}_{-2.1}$&$1,502_{-95}^{+76}$&$0.49_{-0.05}^{+0.04}$&$4.06_{-0.38}^{+0.38}$&$0.38_{-0.04}^{+0.04}$&$0.82_{-0.15}^{+0.15}$\\
A1060&$3.24^{+0.06}_{-0.06}$&0.09&$0.607^{+0.040}_{-0.034}$&$66.2^{+10.9}_{-9.9}$&$790_{-176}^{+171}$&$0.07_{-0.02}^{+0.02}$&$1.69_{-0.14}^{+0.16}$&$0.10_{-0.02}^{+0.02}$&$0.50_{-0.08}^{+0.09}$\\
A1367&$3.55^{+0.08}_{-0.08}$&0.03&$0.695^{+0.035}_{-0.032}$&$269.7^{+20.4}_{-21.7}$&$1,234_{-141}^{+131}$&$0.27_{-0.04}^{+0.04}$&$3.19_{-0.22}^{+0.24}$&$0.26_{-0.03}^{+0.03}$&$0.75_{-0.10}^{+0.11}$\\
MKW4&$1.71^{+0.09}_{-0.09}$&0.57&$0.440^{+0.004}_{-0.005}$&$7.7^{+0.8}_{-0.8}$&$948_{-110}^{+108}$&$0.09_{-0.01}^{+0.01}$&$0.78_{-0.04}^{+0.04}$&$0.05_{-0.00}^{+0.00}$&$0.08_{-0.01}^{+0.01}$\\
ZwCl1215&$5.58^{+0.89}_{-0.78}$&0.05&$0.819^{+0.038}_{-0.034}$&$303.5^{+23.5}_{-24.5}$&$1,485_{-167}^{+155}$&$0.59_{-0.10}^{+0.10}$&$7.15_{-1.08}^{+1.23}$&$0.72_{-0.12}^{+0.13}$&$2.50_{-0.71}^{+0.81}$\\
NGC4636&$0.76^{+0.01}_{-0.01}$&0.33&$0.491^{+0.032}_{-0.027}$&$4.2^{+2.1}_{-1.4}$&$216_{-92}^{+118}$&$.001_{-.001}^{+.001}$&$.088_{-.007}^{+.008}$&$.001_{-.000}^{+.001}$&$.019_{-.003}^{+.003}$\\
A3526&$3.68^{+0.06}_{-0.06}$&0.29&$0.495^{+0.011}_{-0.010}$&$26.1^{+3.7}_{-3.2}$&$1,175_{-174}^{+189}$&$0.20_{-0.04}^{+0.05}$&$2.35_{-0.08}^{+0.08}$&$0.17_{-0.02}^{+0.02}$&$0.45_{-0.03}^{+0.03}$\\
A1644&$4.70^{+0.90}_{-0.70}$&0.04&$0.579^{+0.111}_{-0.074}$&$211.3^{+90.6}_{-65.9}$&$1,830_{-958}^{+937}$&$0.82_{-0.58}^{+0.62}$&$5.39_{-1.26}^{+1.79}$&$0.59_{-0.23}^{+0.28}$&$0.98_{-0.45}^{+0.64}$\\
A1650&$6.70^{+0.80}_{-0.80}$&0.08&$0.704^{+0.131}_{-0.081}$&$197.9^{+73.7}_{-51.2}$&$1,600_{-758}^{+714}$&$0.69_{-0.47}^{+0.52}$&$8.16_{-1.65}^{+2.36}$&$0.85_{-0.31}^{+0.38}$&$2.81_{-1.07}^{+1.53}$\\
A1651&$6.10^{+0.40}_{-0.40}$&0.15&$0.643^{+0.014}_{-0.013}$&$127.5^{+8.9}_{-10.1}$&$1,725_{-168}^{+151}$&$0.81_{-0.13}^{+0.12}$&$7.38_{-0.53}^{+0.53}$&$0.81_{-0.08}^{+0.08}$&$2.03_{-0.28}^{+0.28}$\\
COMA&$8.38^{+0.34}_{-0.34}$&0.06&$0.654^{+0.019}_{-0.021}$&$242.3^{+18.6}_{-20.1}$&$1,954_{-202}^{+201}$&$1.13_{-0.19}^{+0.18}$&$11.57_{-0.70}^{+0.67}$&$1.38_{-0.13}^{+0.13}$&$3.81_{-0.44}^{+0.42}$\\
NGC5044&$1.07^{+0.01}_{-0.01}$&0.67&$0.524^{+0.002}_{-0.003}$&$7.7^{+0.8}_{-0.8}$&$487_{-53}^{+50}$&$.015_{-.003}^{+.002}$&$.300_{-.004}^{+.003}$&$.010_{-.001}^{+.001}$&$.043_{-.001}^{+.001}$\\
A1736&$3.50^{+0.40}_{-0.40}$&0.03&$0.542^{+0.147}_{-0.092}$&$263.4^{+125.8}_{-92.7}$&$1,889_{-1,229}^{+1,110}$&$0.82_{-0.66}^{+0.66}$&$3.85_{-1.03}^{+1.54}$&$0.42_{-0.19}^{+0.23}$&$0.48_{-0.25}^{+0.38}$\\
A3558&$5.50^{+0.40}_{-0.40}$&0.09&$0.580^{+0.006}_{-0.005}$&$157.7^{+7.5}_{-9.6}$&$2,021_{-134}^{+106}$&$1.14_{-0.12}^{+0.10}$&$7.03_{-0.52}^{+0.52}$&$0.84_{-0.07}^{+0.07}$&$1.37_{-0.20}^{+0.20}$\\
A3562&$5.16^{+0.16}_{-0.16}$&0.11&$0.472^{+0.006}_{-0.006}$&$69.7^{+4.6}_{-5.3}$&$1,926_{-167}^{+151}$&$0.83_{-0.10}^{+0.09}$&$5.14_{-0.18}^{+0.18}$&$0.56_{-0.04}^{+0.03}$&$0.81_{-0.06}^{+0.06}$\\
A3571&$6.90^{+0.20}_{-0.20}$&0.14&$0.613^{+0.010}_{-0.010}$&$127.5^{+7.3}_{-8.7}$&$1,897_{-154}^{+137}$&$1.02_{-0.14}^{+0.12}$&$8.76_{-0.32}^{+0.32}$&$1.02_{-0.07}^{+0.06}$&$2.37_{-0.17}^{+0.17}$\\
A1795&$7.80^{+1.00}_{-1.00}$&0.5&$0.596^{+0.003}_{-0.002}$&$54.9^{+2.4}_{-3.2}$&$1,773_{-107}^{+81}$&$0.83_{-0.09}^{+0.07}$&$9.03_{-1.16}^{+1.16}$&$0.99_{-0.14}^{+0.13}$&$2.84_{-0.69}^{+0.69}$\\
A3581&$1.83^{+0.04}_{-0.04}$&0.31&$0.543^{+0.024}_{-0.022}$&$24.6^{+3.7}_{-3.1}$&$840_{-169}^{+174}$&$0.08_{-0.02}^{+0.02}$&$0.92_{-0.06}^{+0.06}$&$0.05_{-0.01}^{+0.01}$&$0.14_{-0.02}^{+0.02}$\\
MKW8&$3.29^{+0.23}_{-0.22}$&0.05&$0.511^{+0.098}_{-0.059}$&$75.4^{+49.4}_{-29.9}$&$977_{-619}^{+703}$&$0.11_{-0.09}^{+0.12}$&$1.79_{-0.32}^{+0.50}$&$0.11_{-0.05}^{+0.06}$&$0.38_{-0.13}^{+0.21}$\\
A2029&$9.10^{+1.00}_{-1.00}$&0.56&$0.582^{+0.004}_{-0.004}$&$58.5^{+2.8}_{-3.6}$&$2,200_{-146}^{+120}$&$1.55_{-0.17}^{+0.14}$&$12.77_{-1.41}^{+1.41}$&$1.66_{-0.20}^{+0.19}$&$3.71_{-0.78}^{+0.78}$\\
A2052&$3.03^{+0.04}_{-0.04}$&0.52&$0.526^{+0.005}_{-0.005}$&$26.1^{+1.8}_{-2.0}$&$1,373_{-119}^{+107}$&$0.34_{-0.04}^{+0.04}$&$2.40_{-0.05}^{+0.05}$&$0.21_{-0.01}^{+0.01}$&$0.35_{-0.01}^{+0.01}$\\
MKW3S&$3.70^{+0.20}_{-0.20}$&0.31&$0.581^{+0.008}_{-0.007}$&$46.5^{+2.9}_{-3.4}$&$1,257_{-108}^{+93}$&$0.29_{-0.04}^{+0.03}$&$2.96_{-0.17}^{+0.17}$&$0.24_{-0.02}^{+0.02}$&$0.62_{-0.07}^{+0.07}$\\
A2065&$5.50^{+0.40}_{-0.40}$&0.04&$1.162^{+0.734}_{-0.282}$&$485.9^{+254.4}_{-133.8}$&$1,302_{-1,048}^{+780}$&$0.49_{-0.49}^{+0.49}$&$8.01_{-2.99}^{+7.20}$&$0.76_{-0.76}^{+0.76}$&$3.83_{-2.53}^{+6.10}$\\
A2063&$3.68^{+0.11}_{-0.11}$&0.12&$0.561^{+0.011}_{-0.011}$&$77.5^{+5.9}_{-6.1}$&$1,343_{-130}^{+127}$&$0.33_{-0.05}^{+0.05}$&$3.03_{-0.12}^{+0.12}$&$0.26_{-0.02}^{+0.02}$&$0.58_{-0.05}^{+0.05}$\\
A2142&$9.70^{+1.50}_{-1.10}$&0.27&$0.591^{+0.006}_{-0.006}$&$108.5^{+6.2}_{-7.4}$&$2,537_{-192}^{+167}$&$2.39_{-0.30}^{+0.26}$&$15.93_{-1.82}^{+2.47}$&$2.32_{-0.29}^{+0.38}$&$4.36_{-0.96}^{+1.30}$\\
15&$4.91^{+0.28}_{-0.28}$&0.03&$0.444^{+0.071}_{-0.046}$&$167.6^{+72.9}_{-46.7}$&$2,360_{-1,201}^{+1,215}$&$1.43_{-0.82}^{+0.91}$&$5.62_{-0.88}^{+1.31}$&$0.71_{-0.21}^{+0.26}$&$0.65_{-0.20}^{+0.30}$\\
A2163&$13.29^{+0.64}_{-0.64}$&0.1&$0.796^{+0.030}_{-0.028}$&$365.5^{+26.7}_{-29.0}$&$2,509_{-273}^{+253}$&$3.07_{-0.51}^{+0.48}$&$28.51_{-1.98}^{+2.05}$&$4.42_{-0.44}^{+0.43}$&$13.19_{-1.63}^{+1.69}$\\
A2199&$4.10^{+0.08}_{-0.08}$&0.16&$0.655^{+0.019}_{-0.021}$&$97.9^{+8.2}_{-8.9}$&$1,300_{-154}^{+153}$&$0.36_{-0.07}^{+0.07}$&$3.81_{-0.19}^{+0.17}$&$0.33_{-0.03}^{+0.03}$&$0.96_{-0.09}^{+0.08}$\\
A2204&$7.21^{+0.25}_{-0.25}$&0.99&$0.597^{+0.008}_{-0.007}$&$47.2^{+2.9}_{-3.4}$&$2,216_{-196}^{+169}$&$1.65_{-0.24}^{+0.21}$&$10.46_{-0.40}^{+0.41}$&$1.38_{-0.10}^{+0.09}$&$2.49_{-0.19}^{+0.19}$\\
A2244&$7.10^{+5.00}_{-2.20}$&0.23&$0.607^{+0.016}_{-0.015}$&$88.7^{+8.6}_{-8.6}$&$1,773_{-222}^{+216}$&$0.84_{-0.17}^{+0.17}$&$8.36_{-2.61}^{+5.89}$&$0.92_{-0.30}^{+0.65}$&$2.45_{-1.46}^{+3.31}$\\
A2256&$6.60^{+0.40}_{-0.40}$&0.05&$0.914^{+0.054}_{-0.047}$&$413.4^{+33.1}_{-34.9}$&$1,684_{-208}^{+189}$&$0.95_{-0.95}^{+0.95}$&$10.51_{-1.00}^{+1.09}$&$1.20_{-0.74}^{+0.85}$&$4.13_{-0.72}^{+0.79}$\\
A2255&$6.87^{+0.20}_{-0.20}$&0.03&$0.797^{+0.033}_{-0.030}$&$417.6^{+30.3}_{-32.6}$&$1,730_{-174}^{+160}$&$0.85_{-0.13}^{+0.12}$&$9.82_{-0.60}^{+0.65}$&$1.08_{-0.10}^{+0.10}$&$3.47_{-0.40}^{+0.43}$\\
A3667&$7.00^{+0.60}_{-0.60}$&0.07&$0.541^{+0.008}_{-0.008}$&$196.5^{+10.9}_{-13.1}$&$2,589_{-199}^{+175}$&$2.24_{-0.25}^{+0.22}$&$10.69_{-0.94}^{+0.94}$&$1.53_{-0.15}^{+0.15}$&$1.93_{-0.34}^{+0.34}$
\end{tabular}
\end{table*}
\clearpage

\setcounter{table}{1}
\begin{table*}
\contcaption{\sc X-ray Cluster Properties of the {\em Complete Sample}:}
\begin{tabular}{lcccrrcccc}
\hline\hline
\colhead{Cluster} & \colhead{$T$} & \colhead{$\rho_{0}$} & \colhead{$\beta$} & \colhead{$r_{c}$} & \colhead{$r_{\rm out}$} & \colhead{$M_{\rm gas}$} & \colhead{$M_{\rm N}$} & \colhead{$M_{\rm MSTG}$} & \colhead{$M_{\rm MOND}$}\\
& \colhead{[keV]} & \colhead{[$10^{-25}\,\mbox{g/cm}^{3}$]} && \colhead{[kpc]} & \colhead{[kpc]} & \colhead{[$10^{14} M_{\sun}$]} & \colhead{[$10^{14} M_{\sun}$]} & \colhead{[$10^{14} M_{\sun}$]} & \colhead{[$10^{14} M_{\sun}$]}\\
\colhead{\footnotesize (1)}&\colhead{\footnotesize (2)}&\colhead{\footnotesize (3)}&\colhead{\footnotesize (4)}&\colhead{\footnotesize (5)}&\colhead{\footnotesize (6)}&\colhead{\footnotesize (7)}&\colhead{\footnotesize (8)}&\colhead{\footnotesize (9)}&\colhead{\footnotesize (10)}\\
\hline
S1101&$3.00^{+1.20}_{-0.70}$&0.55&$0.639^{+0.006}_{-0.007}$&$39.4^{+2.2}_{-2.6}$&$1,064_{-78}^{+70}$&$0.20_{-0.03}^{+0.02}$&$2.23_{-0.52}^{+0.89}$&$0.17_{-0.04}^{+0.07}$&$0.50_{-0.23}^{+0.39}$\\
A2589&$3.70^{+2.20}_{-1.10}$&0.12&$0.596^{+0.013}_{-0.012}$&$83.1^{+6.6}_{-6.8}$&$1,206_{-121}^{+116}$&$0.25_{-0.04}^{+0.04}$&$2.90_{-0.87}^{+1.73}$&$0.23_{-0.07}^{+0.14}$&$0.65_{-0.38}^{+0.76}$\\
A2597&$4.40^{+0.40}_{-0.70}$&0.71&$0.633^{+0.008}_{-0.008}$&$40.8^{+2.2}_{-2.7}$&$1,296_{-103}^{+91}$&$0.36_{-0.05}^{+0.04}$&$3.96_{-0.63}^{+0.37}$&$0.34_{-0.06}^{+0.04}$&$1.04_{-0.32}^{+0.19}$\\
A2634&$3.70^{+0.28}_{-0.28}$&0.02&$0.640^{+0.051}_{-0.043}$&$256.3^{+32.8}_{-31.0}$&$1,225_{-219}^{+208}$&$0.25_{-0.06}^{+0.06}$&$3.05_{-0.37}^{+0.41}$&$0.24_{-0.04}^{+0.04}$&$0.69_{-0.17}^{+0.18}$\\
A2657&$3.70^{+0.30}_{-0.30}$&0.1&$0.556^{+0.008}_{-0.007}$&$83.8^{+5.0}_{-5.9}$&$1,307_{-105}^{+90}$&$0.30_{-0.04}^{+0.03}$&$2.94_{-0.24}^{+0.25}$&$0.24_{-0.02}^{+0.02}$&$0.57_{-0.09}^{+0.09}$\\
A4038&$3.15^{+0.03}_{-0.03}$&0.26&$0.541^{+0.009}_{-0.008}$&$41.5^{+3.3}_{-3.7}$&$1,274_{-134}^{+121}$&$0.28_{-0.04}^{+0.04}$&$2.38_{-0.05}^{+0.06}$&$0.19_{-0.01}^{+0.01}$&$0.40_{-0.02}^{+0.02}$\\
A4059&$4.40^{+0.30}_{-0.30}$&0.2&$0.582^{+0.010}_{-0.010}$&$63.4^{+4.4}_{-5.0}$&$1,324_{-126}^{+116}$&$0.33_{-0.05}^{+0.05}$&$3.71_{-0.27}^{+0.27}$&$0.32_{-0.03}^{+0.03}$&$0.88_{-0.12}^{+0.12}$\\
A2734&$3.85^{+0.62}_{-0.54}$&0.06&$0.624^{+0.034}_{-0.029}$&$149.3^{+19.4}_{-18.3}$&$1,357_{-234}^{+226}$&$0.36_{-0.09}^{+0.09}$&$3.53_{-0.55}^{+0.63}$&$0.31_{-0.06}^{+0.07}$&$0.76_{-0.23}^{+0.27}$\\
A2877&$3.50^{+2.20}_{-1.10}$&0.03&$0.566^{+0.029}_{-0.025}$&$133.8^{+14.5}_{-14.1}$&$943_{-139}^{+132}$&$0.11_{-0.02}^{+0.02}$&$2.01_{-0.64}^{+1.27}$&$0.13_{-0.04}^{+0.08}$&$0.51_{-0.31}^{+0.62}$\\
NGC499&$0.72^{+0.03}_{-0.02}$&0.2&$0.722^{+0.034}_{-0.030}$&$16.9^{+1.6}_{-1.7}$&$196_{-30}^{+27}$&$.001_{-.000}^{+.000}$&$.111_{-.007}^{+.009}$&$.002_{-.000}^{+.000}$&$.035_{-.004}^{+.005}$\\
AWM7&$3.75^{+0.09}_{-0.09}$&0.09&$0.671^{+0.027}_{-0.025}$&$121.8^{+13.7}_{-12.6}$&$1,120_{-154}^{+157}$&$0.22_{-0.05}^{+0.05}$&$3.05_{-0.18}^{+0.19}$&$0.23_{-0.03}^{+0.03}$&$0.83_{-0.09}^{+0.10}$\\
PERSEUS&$6.79^{+0.12}_{-0.12}$&0.63&$0.540^{+0.006}_{-0.004}$&$45.1^{+2.4}_{-2.9}$&$2,414_{-189}^{+145}$&$1.88_{-0.22}^{+0.18}$&$9.70_{-0.20}^{+0.23}$&$1.33_{-0.07}^{+0.06}$&$1.83_{-0.07}^{+0.08}$\\
S405&$4.21^{+0.67}_{-0.59}$&0.02&$0.664^{+0.263}_{-0.133}$&$323.2^{+185.0}_{-113.4}$&$1,561_{-1,165}^{+1,034}$&$0.53_{-0.46}^{+0.53}$&$4.59_{-1.47}^{+2.68}$&$0.44_{-0.22}^{+0.33}$&$0.97_{-0.61}^{+1.11}$\\
3C129&$5.60^{+0.70}_{-0.60}$&0.03&$0.601^{+0.260}_{-0.131}$&$223.9^{+125.7}_{-76.4}$&$1,567_{-1,455}^{+1,113}$&$0.52_{-0.52}^{+0.52}$&$5.67_{-1.85}^{+3.54}$&$0.55_{-0.34}^{+0.43}$&$1.46_{-0.92}^{+1.76}$\\
A0539&$3.24^{+0.09}_{-0.09}$&0.06&$0.561^{+0.020}_{-0.018}$&$104.2^{+10.2}_{-10.3}$&$1,194_{-158}^{+150}$&$0.23_{-0.04}^{+0.04}$&$2.36_{-0.13}^{+0.14}$&$0.18_{-0.02}^{+0.02}$&$0.44_{-0.05}^{+0.05}$\\
S540&$2.40^{+0.38}_{-0.34}$&0.08&$0.641^{+0.073}_{-0.051}$&$91.5^{+27.0}_{-21.1}$&$877_{-305}^{+305}$&$0.10_{-0.05}^{+0.06}$&$1.46_{-0.27}^{+0.33}$&$0.09_{-0.03}^{+0.03}$&$0.31_{-0.11}^{+0.14}$\\
A0548w&$1.20^{+0.19}_{-0.17}$&0.02&$0.666^{+0.194}_{-0.111}$&$139.4^{+63.7}_{-44.4}$&$593_{-328}^{+311}$&$.028_{-.018}^{+.022}$&$.492_{-.137}^{+.218}$&$.021_{-.009}^{+.012}$&$.078_{-.043}^{+.069}$\\
A0548&$3.10^{+0.10}_{-0.10}$&0.05&$0.480^{+0.013}_{-0.013}$&$83.1^{+9.2}_{-9.1}$&$1,324_{-175}^{+177}$&$0.27_{-0.05}^{+0.05}$&$2.15_{-0.11}^{+0.11}$&$0.17_{-0.02}^{+0.02}$&$0.30_{-0.03}^{+0.03}$\\
A3395&$5.00^{+0.30}_{-0.30}$&0.02&$0.981^{+0.619}_{-0.244}$&$473.2^{+270.5}_{-145.4}$&$1,221_{-977}^{+783}$&$0.30_{-0.30}^{+0.30}$&$5.71_{-2.21}^{+5.13}$&$0.48_{-0.48}^{+0.48}$&$2.31_{-1.64}^{+3.80}$\\
UGC03957&$2.58^{+0.41}_{-0.36}$&0.09&$0.740^{+0.133}_{-0.086}$&$100.0^{+32.0}_{-23.9}$&$764_{-339}^{+306}$&$0.08_{-0.05}^{+0.05}$&$1.57_{-0.34}^{+0.47}$&$0.09_{-0.03}^{+0.04}$&$0.47_{-0.19}^{+0.27}$\\
PKS0745&$7.21^{+0.11}_{-0.11}$&0.97&$0.608^{+0.006}_{-0.006}$&$50.0^{+2.5}_{-3.1}$&$2,169_{-159}^{+137}$&$1.59_{-0.20}^{+0.17}$&$10.43_{-0.22}^{+0.22}$&$1.36_{-0.08}^{+0.07}$&$2.58_{-0.10}^{+0.10}$\\
A0644&$7.90^{+0.80}_{-0.80}$&0.15&$0.700^{+0.011}_{-0.011}$&$143.0^{+7.8}_{-9.4}$&$1,557_{-119}^{+103}$&$0.66_{-0.09}^{+0.08}$&$9.37_{-0.97}^{+0.97}$&$0.97_{-0.12}^{+0.11}$&$3.81_{-0.72}^{+0.72}$\\
S636&$1.18^{+0.19}_{-0.17}$&0.01&$0.752^{+0.217}_{-0.123}$&$242.3^{+92.1}_{-62.1}$&$742_{-336}^{+323}$&$0.06_{-0.02}^{+0.04}$&$0.65_{-0.18}^{+0.29}$&$0.03_{-0.01}^{+0.02}$&$0.09_{-0.05}^{+0.08}$\\
A1413&$7.32^{+0.26}_{-0.24}$&0.19&$0.660^{+0.017}_{-0.015}$&$126.1^{+10.0}_{-10.5}$&$1,794_{-194}^{+179}$&$0.95_{-0.17}^{+0.16}$&$9.46_{-0.43}^{+0.48}$&$1.08_{-0.10}^{+0.10}$&$3.03_{-0.26}^{+0.29}$\\
M49&$0.95^{+0.02}_{-0.01}$&0.26&$0.592^{+0.007}_{-0.007}$&$7.7^{+0.8}_{-0.8}$&$177_{-20}^{+19}$&$.001_{-.000}^{+.000}$&$.109_{-.002}^{+.003}$&$.002_{-.000}^{+.000}$&$.041_{-.001}^{+.002}$\\
A3528n&$3.40^{+1.66}_{-0.64}$&0.07&$0.621^{+0.034}_{-0.030}$&$125.4^{+13.1}_{-13.3}$&$1,181_{-193}^{+179}$&$0.24_{-0.06}^{+0.05}$&$2.71_{-0.54}^{+1.34}$&$0.21_{-0.05}^{+0.11}$&$0.59_{-0.23}^{+0.57}$\\
A3528s&$3.15^{+0.89}_{-0.59}$&0.09&$0.463^{+0.013}_{-0.012}$&$71.1^{+7.0}_{-6.9}$&$1,872_{-250}^{+244}$&$0.75_{-0.13}^{+0.12}$&$2.99_{-0.57}^{+0.85}$&$0.32_{-0.07}^{+0.09}$&$0.29_{-0.11}^{+0.17}$\\
A3530&$3.89^{+0.27}_{-0.25}$&0.03&$0.773^{+0.114}_{-0.085}$&$296.5^{+54.3}_{-46.1}$&$1,150_{-308}^{+282}$&$0.23_{-0.07}^{+0.08}$&$3.56_{-0.60}^{+0.79}$&$0.28_{-0.06}^{+0.07}$&$1.05_{-0.34}^{+0.44}$\\
A3532&$4.58^{+0.19}_{-0.17}$&0.05&$0.653^{+0.034}_{-0.029}$&$198.6^{+20.8}_{-20.3}$&$1,372_{-199}^{+188}$&$0.38_{-0.08}^{+0.08}$&$4.41_{-0.32}^{+0.37}$&$0.39_{-0.05}^{+0.05}$&$1.15_{-0.16}^{+0.19}$\\
A1689&$9.23^{+0.28}_{-0.28}$&0.33&$0.690^{+0.011}_{-0.011}$&$114.8^{+6.9}_{-7.7}$&$1,898_{-154}^{+143}$&$1.23_{-0.18}^{+0.17}$&$13.21_{-0.50}^{+0.50}$&$1.61_{-0.12}^{+0.11}$&$5.14_{-0.36}^{+0.36}$\\
A3560&$3.16^{+0.51}_{-0.44}$&0.03&$0.566^{+0.033}_{-0.029}$&$180.3^{+22.5}_{-21.6}$&$1,402_{-240}^{+230}$&$0.35_{-0.08}^{+0.08}$&$2.71_{-0.42}^{+0.49}$&$0.23_{-0.04}^{+0.05}$&$0.42_{-0.13}^{+0.15}$\\
A1775&$3.69^{+0.20}_{-0.11}$&0.06&$0.673^{+0.026}_{-0.023}$&$183.1^{+15.5}_{-16.3}$&$1,391_{-167}^{+153}$&$0.42_{-0.08}^{+0.07}$&$3.72_{-0.21}^{+0.29}$&$0.34_{-0.03}^{+0.04}$&$0.81_{-0.09}^{+0.12}$\\
A1800&$4.02^{+0.64}_{-0.56}$&0.04&$0.766^{+0.308}_{-0.139}$&$276.1^{+157.5}_{-94.2}$&$1,284_{-949}^{+825}$&$0.34_{-0.23}^{+0.36}$&$4.14_{-1.23}^{+2.45}$&$0.36_{-0.15}^{+0.27}$&$1.15_{-0.66}^{+1.31}$\\
A1914&$10.53^{+0.51}_{-0.50}$&0.22&$0.751^{+0.018}_{-0.017}$&$162.7^{+10.4}_{-11.6}$&$1,768_{-162}^{+148}$&$1.08_{-0.18}^{+0.16}$&$15.21_{-0.87}^{+0.90}$&$1.79_{-0.16}^{+0.16}$&$7.44_{-0.75}^{+0.77}$\\
NGC5813&$0.52^{+0.08}_{-0.07}$&0.18&$0.766^{+0.179}_{-0.103}$&$17.6^{+6.4}_{-4.3}$&$166_{-97}^{+79}$&$.001_{-.001}^{+.001}$&$.072_{-.017}^{+.026}$&$.001_{-.000}^{+.001}$&$.021_{-.009}^{+.014}$\\
NGC5846&$0.82^{+0.01}_{-0.01}$&0.47&$0.599^{+0.016}_{-0.015}$&$4.9^{+0.7}_{-0.8}$&$152_{-27}^{+26}$&$.001_{-.000}^{+.000}$&$.082_{-.003}^{+.003}$&$.001_{-.000}^{+.000}$&$.031_{-.002}^{+.002}$\\
A2151w&$2.40^{+0.06}_{-0.06}$&0.16&$0.564^{+0.014}_{-0.013}$&$47.9^{+4.1}_{-4.4}$&$957_{-114}^{+105}$&$0.12_{-0.02}^{+0.02}$&$1.42_{-0.06}^{+0.06}$&$0.09_{-0.01}^{+0.01}$&$0.25_{-0.02}^{+0.02}$\\
A3627&$6.02^{+0.08}_{-0.08}$&0.04&$0.555^{+0.056}_{-0.044}$&$210.6^{+40.4}_{-36.5}$&$1,830_{-515}^{+474}$&$0.78_{-0.29}^{+0.28}$&$6.62_{-0.75}^{+0.95}$&$0.71_{-0.14}^{+0.15}$&$1.47_{-0.32}^{+0.41}$\\
TRIANGUL&$9.60^{+0.60}_{-0.60}$&0.1&$0.610^{+0.010}_{-0.010}$&$196.5^{+11.4}_{-13.1}$&$2,385_{-187}^{+169}$&$1.98_{-0.25}^{+0.22}$&$15.22_{-1.01}^{+1.01}$&$2.11_{-0.18}^{+0.17}$&$4.48_{-0.57}^{+0.57}$\\
OPHIUCHU&$10.26^{+0.32}_{-0.32}$&0.13&$0.747^{+0.035}_{-0.032}$&$196.5^{+18.2}_{-19.0}$&$1,701_{-240}^{+224}$&$0.91_{-0.21}^{+0.20}$&$14.11_{-0.96}^{+1.03}$&$1.59_{-0.19}^{+0.19}$&$6.91_{-0.83}^{+0.89}$\\
ZwC174&$5.23^{+0.84}_{-0.73}$&0.1&$0.717^{+0.073}_{-0.053}$&$163.4^{+33.1}_{-28.3}$&$1,354_{-378}^{+349}$&$0.43_{-0.18}^{+0.19}$&$5.49_{-0.96}^{+1.18}$&$0.50_{-0.13}^{+0.15}$&$1.79_{-0.59}^{+0.73}$\\
A2319&$8.80^{+0.50}_{-0.50}$&0.1&$0.591^{+0.013}_{-0.012}$&$200.7^{+13.5}_{-15.0}$&$2,657_{-250}^{+228}$&$2.66_{-0.39}^{+0.35}$&$15.07_{-0.96}^{+0.98}$&$2.25_{-0.20}^{+0.19}$&$3.60_{-0.45}^{+0.45}$\\
A3695&$5.29^{+0.85}_{-0.74}$&0.04&$0.642^{+0.259}_{-0.117}$&$281.0^{+179.3}_{-106.1}$&$1,887_{-1,652}^{+1,379}$&$0.96_{-0.96}^{+0.96}$&$6.88_{-2.03}^{+4.08}$&$0.79_{-0.46}^{+0.62}$&$1.49_{-0.86}^{+1.73}$\\
IIZw108&$3.44^{+0.55}_{-0.48}$&0.03&$0.662^{+0.167}_{-0.097}$&$257.0^{+112.5}_{-75.3}$&$1,327_{-695}^{+670}$&$0.33_{-0.22}^{+0.26}$&$3.19_{-0.80}^{+1.25}$&$0.27_{-0.11}^{+0.14}$&$0.65_{-0.32}^{+0.50}$\\
A3822&$4.90^{+0.78}_{-0.69}$&0.04&$0.639^{+0.150}_{-0.093}$&$247.2^{+113.2}_{-79.4}$&$1,675_{-942}^{+904}$&$0.67_{-0.52}^{+0.57}$&$5.63_{-1.41}^{+2.07}$&$0.58_{-0.25}^{+0.31}$&$1.27_{-0.62}^{+0.91}$\\
A3827&$7.08^{+1.13}_{-0.99}$&0.05&$0.989^{+0.410}_{-0.192}$&$417.6^{+175.5}_{-107.5}$&$1,515_{-977}^{+762}$&$0.74_{-0.74}^{+0.74}$&$10.81_{-3.39}^{+6.58}$&$1.15_{-1.15}^{+1.15}$&$5.16_{-2.87}^{+5.57}$\\
A3888&$8.84^{+1.41}_{-1.24}$&0.1&$0.928^{+0.084}_{-0.066}$&$282.4^{+34.5}_{-32.4}$&$1,455_{-281}^{+252}$&$0.71_{-0.71}^{+0.71}$&$12.61_{-2.18}^{+2.58}$&$1.33_{-1.33}^{+1.33}$&$7.14_{-2.07}^{+2.45}$\\
A3921&$5.73^{+0.24}_{-0.23}$&0.07&$0.762^{+0.036}_{-0.030}$&$231.0^{+20.7}_{-20.8}$&$1,435_{-182}^{+167}$&$0.52_{-0.10}^{+0.10}$&$6.70_{-0.46}^{+0.53}$&$0.65_{-0.07}^{+0.07}$&$2.35_{-0.30}^{+0.35}$\\
HCG94&$3.45^{+0.30}_{-0.30}$&0.11&$0.514^{+0.007}_{-0.006}$&$60.6^{+3.8}_{-4.4}$&$1,237_{-104}^{+89}$&$0.24_{-0.03}^{+0.02}$&$2.40_{-0.21}^{+0.21}$&$0.19_{-0.02}^{+0.02}$&$0.43_{-0.07}^{+0.07}$\\
RXJ2344&$4.73^{+0.76}_{-0.66}$&0.07&$0.807^{+0.033}_{-0.030}$&$212.0^{+16.7}_{-17.4}$&$1,222_{-135}^{+127}$&$0.34_{-0.06}^{+0.05}$&$4.97_{-0.74}^{+0.85}$&$0.43_{-0.07}^{+0.08}$&$1.78_{-0.50}^{+0.57}$\\
\hline
\end{tabular}
\begin{tabular}{llll}
\multicolumn{4}{l}{Note --- We have adopted the compilation of \citet{rei01, rei02} as our sample.}\\
Column (1)&Galaxy cluster name (truncated to 8 characters)&Column (6)&radius where $\rho_{gas} \approx 10^{-28}\,\mbox{g/cm}^{3}$\\
Column (2)&X-ray temperature&Column (7)&ICM gas mass integrated to $r_{\rm out}$\\
Column (3)&ICM central mass density&Column (8)&Newtonian dynamic mass integrated to $r_{\rm out}$\\
Column (4)&$\beta$-model $\beta$ parameter&Column (9)&MSTG dynamic mass integrated to $r_{\rm out}$\\
Column (5)&$\beta$-model core radius parameter&Column (10)&convergent MOND dynamic mass
\end{tabular}
\end{table*}
\label{lastpage}
\clearpage


\begin{thebibliography}{}
\addcontentsline{toc}{section}{References}
\bibitem[\protect\citeauthoryear{Aguirre \etal}{2001}]{agu01} Aguirre, A., Schaye, J. \& Quataert, E. 2001, \apj, 561, 550 
\bibitem[\protect\citeauthoryear{Arnaud \etal}{2001a}]{arn01a} Arnaud, M. \etal 2001,  \aap, 365, L67 
\bibitem[\protect\citeauthoryear{Arnaud \etal}{2001b}]{arn01b} Arnaud, M., Neumann D.\,M., Aghanim, N., Gastaud, R., Majerowicz, S. \& Hughes, J.\,P. 2001, \aap, 365, L80 
\bibitem[\protect\citeauthoryear{Brownstein \& Moffat}{2005}]{bro05} Brownstein, J.\,R. \& Moffat, J.\,W. 2005, \apj, submitted (\href{http://arxiv.org/abs/astro-ph/0506370}{\tt astro-ph/0506370})
\bibitem[\protect\citeauthoryear{Cavaliere \& Fusco-Femiano}{1976}]{cav76} Cavaliere, A.\,L. \& Fusco-Femiano, R. 1976, \aap, 49, 137
\bibitem[\protect\citeauthoryear{Chandrasekhar}{1960}]{cha60} Chandrasekhar, S. 1960, {\em Principles of Stellar Dynamics}, Dover, New York
\bibitem[\protect\citeauthoryear{Eidelman \etal}{2004}]{eid04} Eidelman, S. \etal 2004, Phys. Lett. B592, 1,  2005 update (\href{http://pdg.lbl.gov}{\tt http://pdg.lbl.gov})
\bibitem[\protect\citeauthoryear{King}{1966}]{kin66} King, I.\,R. 1966, \aj, 71, 64
\bibitem[\protect\citeauthoryear{Milgrom}{1983}]{mil83} Milgrom, M. 1983, \apj, 270, 365
\bibitem[\protect\citeauthoryear{Milgrom}{1984}]{mil84} Milgrom, M. 1983, \apj, 287, 571
\bibitem[\protect\citeauthoryear{Moffat}{2005a}]{mof05a}  Moffat, J.\,W. 2005, \jcap, 05, 003 
\bibitem[\protect\citeauthoryear{Moffat}{2005b}]{mof05b} Moffat, J.\,W. 2005, preprint (\href{http://arxiv.org/abs/gr-qc/0506021}{\tt gr-qc/0506021})
\bibitem[\protect\citeauthoryear{Navarro \etal}{1997}]{nav97} Navarro, J.\,F., Frenk, C.\,S., \& White, S.\,D.\,M. 1997, \apj, 490, 493  
\bibitem[\protect\citeauthoryear{Ota \& Mitsuda}{2002}]{ota02} Ota, N. \& Mitsuda, K. 2002, \apj, 567, L23 
\bibitem[\protect\citeauthoryear{Pointecouteau \etal}{2005}]{poi05a} Pointecouteau, E., Arnaud, M. \& Pratt, G.\,W. 2005, \aap, 435, 1 
\bibitem[\protect\citeauthoryear{Pointecouteau \& Silk}{2005}]{poi05b} Pointecouteau, E. \& Silk, J. 2005, \mnras, submitted (\href{http://arxiv.org/abs/astro-ph/0505017}{\tt astro-ph/0505017})
\bibitem[\protect\citeauthoryear{Prada \etal}{2003}]{pra03} Prada, F. \etal 2003, \apj, 598, 260 
\bibitem[\protect\citeauthoryear{Reiprich}{2001}]{rei01} Reiprich, T.\,H. 2001, Ph.D.\,Dissertation, {\em Cosmological Implications and Physical Properties of an X-Ray Flux-Limited Sample of Galaxy Clusters}, Ludwig-Maximilians-Universit\"at M\"unchen 
\bibitem[\protect\citeauthoryear{Reiprich \& B\"ohringer}{2002}]{rei02} Reiprich, T.\,H. \& B\"ohringer, H. 2002, \apj,  567, 716 
\bibitem[\protect\citeauthoryear{Reuter \& Weyer}{2004a}]{reu04a} Reuter, M. \& Weyer, H. 2004, \jcap, 12, 001 
\bibitem[\protect\citeauthoryear{Reuter \& Weyer}{2004b}]{reu04b} Reuter, M. \& Weyer, H. 2004, \prd 69, 104022 
\bibitem[\protect\citeauthoryear{Sanders}{2003}]{san03} Sanders, R.\,H. 2003, \mnras, 342, 901 
\bibitem[\protect\citeauthoryear{Sanders \& McGaugh}{2002}]{san02} Sanders, R.\,H. \& McGaugh, S.\,S. 2002, \araa 40, 263 
\bibitem[\protect\citeauthoryear{The \& White}{1988}]{the88} The L.\,S. \& White S,\,D.\,M. 1988, \aj, 95, 1642
\bibitem[\protect\citeauthoryear{Zwicky}{1933}]{zwi33} Zwicky, F. 1933, Helv.\,Phys.\,Acta., 6, 110
\end{thebibliography}
\end{document}